\documentclass[sigconf]{acmart}

\AtBeginDocument{%
  }

\copyrightyear{2025}
\acmYear{2025}
\setcopyright{cc}
\setcctype{by}
\acmConference[UMAP Adjunct '25]{Adjunct Proceedings of the 33rd ACM Conference on User Modeling, Adaptation and Personalization}{June 16--19, 2025}{New York City, NY, USA}
\acmBooktitle{Adjunct Proceedings of the 33rd ACM Conference on User Modeling, Adaptation and Personalization (UMAP Adjunct '25), June 16--19, 2025, New York City, NY, USA}
\acmDOI{10.1145/3708319.3733710}
\acmISBN{979-8-4007-1399-6/2025/06}

\usepackage{graphicx}
\usepackage{subcaption}

\begin{document}

\title{User and Recommender Behavior Over Time}\subtitle{Contextualizing Activity, Effectiveness, Diversity, and Fairness in Book Recommendation}

\author{Samira Vaez Barenji}
\email{sv849@drexel.edu}
\orcid{0000-0002-2123-4338}
\affiliation{%
  \institution{Dept. of Information Science \\ Drexel University}
  \city{Philadelphia}
  \state{PA}
  \country{USA}
}
\author{Sushobhan Parajuli}
\email{sp3886@drexel.edu}
\orcid{0009-0008-5679-9524}
\affiliation{%
  \institution{Dept. of Information Science \\ Drexel University}
  \city{Philadelphia}
  \state{PA}
  \country{USA}
}
\author{Michael D. Ekstrand}
\email{mdekstrand@drexel.edu}
\orcid{0000-0003-2467-0108}
\affiliation{%
  \institution{Dept. of Information Science \\ Drexel University}
  \city{Philadelphia}
  \state{PA}
  \country{USA}
}

\renewcommand{\shortauthors}{Vaez Barenji et al.}

\begin{abstract}
Data is an essential resource for studying recommender systems. While there has been significant work on improving and evaluating state-of-the-art models and measuring various properties of recommender system outputs, less attention has been given to the data itself, particularly how data has changed over time.
Such documentation and analysis provide guidance and context for designing and evaluating recommender systems, particularly for evaluation designs making use of time (e.g., temporal splitting).  
In this paper, we present a temporal explanatory analysis of the UCSD Book Graph dataset scraped from Goodreads, a social reading and recommendation platform active since 2006. 
We measure the book interaction data using a set of activity, diversity, and fairness metrics; we then train a set of collaborative filtering algorithms on rolling training windows to observe how the same measures evolve over time in the recommendations.
Additionally, we explore whether the introduction of algorithmic recommendations in 2011 was followed by observable changes in user or recommender system behavior.
\end{abstract}





\maketitle

\section{Introduction}
Datasets are crucial to building, evaluating, and studying recommender systems and other information access systems (IAS).
However, the documentation and descriptive analysis accompanying most datasets is slim, focused primarily on high-level descriptive statistics (e.g., the numbers of users, items, and interactions).
As recommender systems research has paid increasing attention to time and sequence in training and evaluation data, both through techniques like sequential recommendation \cite{boka2024survey} and the growing consensus towards global temporal splitting for evaluating even traditional recommender systems \cite{meng2020exploring}, it is important to understand not only the \emph{static} characteristics of a dataset but also how those characteristics \emph{change} over the history of interactions recorded in the dataset. 
Temporal understanding is also valuable for documenting and contextualizing changes in user or system behavior, both for traditional evaluation and to support research on evolving concerns regarding fairness, isolation, and other social impacts.

In this paper, we present a longitudinal explanatory analysis of the UCSD Book Graph \cite{wan2018item, wan2019fine}, a dataset containing book metadata, reviews, ratings, and interactions from the Goodreads social reading and recommendation platform, spanning more than 10 years from early 2007 to 2017.
This analysis describes how the data has evolved over time in its volume as well as genre diversity and gender balance in user interactions, and the effectiveness, diversity, and fairness of collaborative filtering models trained and evaluated at different points in its history.

For the first several years, Goodreads' recommendations were purely social (direct, personal recommendations between users in the social graph).
In 2011, Goodreads introduced algorithmic recommendations, generating recommendations for users after they rated at least 20 items on a five-star scale \cite{goodreads2011}.
We further examine whether there are measurable changes in user or recommender behavior on the dimensions we are analyzing associated with the introduction of this recommender system.

Describing data, user, and model changes over time is important for several reasons: 
i) guiding the design and evaluation of experimental setups or understanding the impact of design decisions (e.g., the effect of selecting different time points for splitting); 
ii) revealing how different types of bias evolve over time, both to directly understand fairness implications of the data as well as to provide temporal context for further fairness investigations; 
iii) assessing whether model performance and fairness measures remain consistent across time. 
While some prior studies have conducted exploratory analyses of recommendation datasets in different domains \cite{thelwall2017goodreads, mishra2019big}, there remains a gap in integrating longitudinal analyses of real-world recommendation datasets into fairness literature. 
This paper addresses this gap through the following  contributions:
\begin{enumerate}
\item Conducting a longitudinal exploratory analysis of user activity on Goodreads to understand how activity, fairness, and diversity metrics of user interactions evolve over time.
\item Comparing four collaborative filtering algorithms trained on successive time windows to examine how their behavior changes over time in relation to the same set of concerns.
\item Investigation of whether the introduction of a recommender system on Goodreads is associated with measurable changes in the system's behavior
\end{enumerate}

The temporal changes in data and evaluation metrics we document show that temporal recommender system evaluations are sensitive to the time window at which they are performed.
Therefore, evaluating a system at a single point in time does not necessarily reflect its broader behavior.
Some periods may reflect peak performance, while others reveal reduced effectiveness or increased disparities. 
Without a temporal perspective, evaluations risk overlooking how biases evolve as both the user base and content shift. 
Understanding system behavior over time is therefore essential for capturing its ability to meet users’ needs and for identifying when and how bias emerges or is mitigated. 
It also supports more meaningful metric interpretation and the development of emerging best practices.

In the remainder of this paper, we survey related work on dataset description, bias, and longitudinal analysis (\S\ref{sec:background}), detail the data, algorithms, and experimental setup (\S\ref{sec:methods}), present our results and observations (\S\ref{sec:results}), and conclude with implications for recommender systems research on fairness and other topics (\S\ref{sec:conclusion}).

\section{Background and Related Work}
\label{sec:background}
\paragraph{Describing Datasets}
There is a growing call for detailed documentation of AI training and evaluation data, motivated by concerns about transparency, accountability, and the potential for unintended harms \citep{gebru2021datasheets, pushkarna2022datacards, nesvijevskaia2021databook, giner2022describeml}. 
Poorly documented datasets can obscure hidden biases, mislead users about appropriate applications, or lead to flawed evaluations \citep{olteanu2019social, pagano2023bias}. 
To address these issues, several documentation frameworks have been proposed. 
\citet{gebru2021datasheets} introduce \textit{datasheets} to describe the dataset's context, content, and recommended uses, similar to documentation practices in engineering. \citet{pushkarna2022datacards} propose \textit{data cards}, which are short summaries that explain how the dataset was created and how it should be used. \citet{nesvijevskaia2021databook} presents a standardized framework \textit{databook} for dynamically documenting algorithm design throughout data science projects.

In the context of recommender systems, the widely-used MovieLens datasets have detailed historical documentation \cite{harper2015movielens}, but this documentation was only published after the platform had been operating for over 15 years. 
However, such documentation is rare for other datasets. Comprehensive data documentation has been identified as essential to enable reproducible results in AI research \cite{gundersen2018state}.
In this paper, we examine the UCSD Book Graph scraped from Goodreads. 
The dataset is briefly summarized on its website\footnote{\url{https://cseweb.ucsd.edu/~jmcauley/datasets/goodreads.html}} and in the authors’ papers \citep{wan2018item, wan2019fine}. 
\citet{ekstrand2021exploring} augmented the Book Graph with publicly available book and author metadata to study fairness towards authors' gender identities; they provided a brief description of the distributions of author gender data. 
However, changes to the dataset's size, composition, and distributions over time have not yet been documented.

\paragraph{Fairness and Social Impacts}
User-generated data on social media platforms and review sites often reflect societal biases. 
These biases can enter datasets through uneven participation, visibility, or representation, and may be further amplified during data processing \cite{olteanu2019social}. 
When such data are used to train recommender systems, it can reinforce existing patterns of popularity or exclusion and lead to biased system behavior \citep{abdollahpouri2017popularitybias, yalcin2022evaluating, beutel2019fairness}. 

Research on fairness in recommender systems seeks to document, measure, and mitigate these biases \citep{ekstrand2022fairness, wang2023surveyfairness, deldjoo2024fairness}. 
\citet{ekstrand2021exploring} used the Book Graph we examine in this paper to study author gender representation in user profiles and recommender system outputs.
Although much of the recommender systems fairness research has focused on static evaluations \cite{chenBiasDebiasRecommender2023}, recent studies highlight the importance of understanding how fairness outcomes evolve as models are retrained on new user feedback \citep{ferraro2024itsnotyou, ferraro2021breaktheloop}.
Much of this work, however, is prospective, examining how fairness changes as models are trained and retrained going forward from a static dataset.
Our present study complements this direction of research by studying and evaluating how user interactions and model behavior evolve historically over time in the Book Graph.
 
\paragraph{Longitudinal Analysis}
Recommender systems can have long-term effects on user behavior and content exposure \cite{ge2021towards}. 
Over time, they may reinforce existing patterns, leading to reduced diversity and unequal visibility across groups \cite{fabbri2022exposure}. 
Some studies have shown that short-term engagement signals do not always align with long-term outcomes, like user satisfaction or return visits \cite{wang2022surrogate}. 
Longitudinal studies, which track user interactions over extended periods, are essential to understanding long-term effects. Many longitudinal studies of recommender systems are prospective and examine how systems and user behavior evolve going forward.
Some, however, are historical, such as the filter bubble analysis of \citet{nguyenExploringFilterBubble2014} and temporal evaluations of recommender system accuracy \citep{burkeEvaluatingDynamicProperties2010, lathiaEvaluatingCollaborativeFiltering2009}.
In this study, we conduct a historical longitudinal analysis to understand how the underlying data in a real-world recommendation platform has changed over time.

\section{Data and Methods}
\label{sec:methods}
To document the evolution of book interaction data and its influence on collaborative filtering outputs, we measure the volume and diversity of user interactions across successive time windows, as well as the results of training and evaluating collaborative filtering models at different points in the dataset’s temporal span. 
This section details the specific methods and data preparation used in our analysis.\footnote{The code is available at \url{https://zenodo.org/records/15333266}}

\subsection{Data}
The UCSD Book Graph \cite{wan2018item, wan2019fine} is a dataset scraped in 2018 from Goodreads, a social reading and recommendation platform. 
It spans over a decade of book metadata, reviews, ratings, and interactions of users with public profiles, covering the period from January 2007 to November 2017.
\citet{ekstrand2021exploring} augmented the book graph with public author data from library sources and OpenLibrary through the PIReT/INERTIAL Book Data Tools\footnote{\url{https://bookdata.inertial.science}}, providing further information such as author gender identities. 
The original book graph includes interactions of approximately 876K users with 1.52M million books. 
It spans various user activities and is organized into the following components:

\begin{itemize}
    \item User actions, where a user adds a book to one of their “shelves”;
    \item User ratings, provided on a 5-star scale;
    \item Book metadata, including title, genres, author names, and other attributes.
\end{itemize}

For our analysis, we count Goodreads \textit{works} as items for user interaction and recommendation.
A work is the fundamental unit of browsing and cataloging in Goodreads, and represents a particular literary work including its various editions, printings, translations, etc.; individual editions are represented as \textit{books}.
Using version 3.0 of the Book Data Tools, we aggregated and deduplicated these individual actions to form an interaction matrix representing each observed user-work pair with a single record. 
Each record includes the first and last timestamps of any interaction between the user and that work, whether through shelving or rating, along with the last rating the user provided for that work. For this study, we used the last interaction timestamp, as the first timestamps had some erroneous values that made them unreliable for analysis.

To support the analysis of genre diversity we used the book genre information supplied with the Book Graph (and extracted from users' genre-related shelf names) to derive a distribution over genres for each work.
The book graph genre records report the number of times each genre shelf has been applied to a book by users. 
We normalized these counts to obtain $P(g \mid b)$, the probability that book $b$ belongs to genre $g$. 
The book graph also contains textual reviews, but we did not consider those in this analysis. 
Table~\ref{tab:gender_summary} presents a statistical summary of the dataset, including the number of unique users, books, and authors, as well as the average rating in the ratings data. 
The gender-specific columns report these same statistics separately for male and female authors based on available gender data. 
For all experiments and analyses, we excluded authors with ambiguous or missing gender labels (using the same logic as \citet{ekstrand2021exploring}).

\begin{table}
\caption{Summary statistics}
\label{tab:gender_summary}
\begin{tabular}{lrrr}
\toprule
 & Overall & Male & Female \\
\midrule
Unique Users & 876,145 & --- & --- \\
Unique Books & 1,522,486 & 348,302 & 231,345 \\
Unique Authors & 612,241 & 155,888 & 87,904 \\
Average Rating & 3.85 & 3.80 & 3.79 \\
\bottomrule
\end{tabular}
\end{table}

\subsection{Experimental Setup}
We used a subset of actions data for our experiments that ranged from January 1, 2007, to October 31, 2017.
We begin by analyzing the interaction data within this period on a monthly basis to examine its temporal trends. 

To generate and evaluate recommendations, we created successive 2-month windows of test data beginning on Jan. 1, 2009, training the recommender system on the preceding 2 years of interactions and generating top-100 recommendation lists for each user in the test window who also has interactions in the training window.
This setup allows us to analyze how model performance evolves over time.

We used LensKit \cite{ekstrandLensKitPythonNextGeneration2020} to generate and evaluate recommendations from collaborative filtering models trained on these successive training sets. 
Following our prior work using this dataset \citep{ekstrand2021exploring, gridlayout2024amifaraj}, we used three implicit-feedback collaborative filtering algorithms: item-based $k$-NN \citep[ItemKNN, ][]{deshpandeItembasedTopRecommendation2004}, implicit-feedback ALS Matrix Factorization \citep[ImplicitMF, ][]{takacsApplicationsConjugateGradient2011}, and Bayesian Personalized Ranking \citep[BPR, ][]{bpr2009}; we omit User-based $k$-NN due to its computational cost. 
We also considered non-personalized most-popular recommendations (MostPop). 
This selection helps ensure consistency and alignment with our previous studies, extending that line of inquiry through a longitudinal lens.

\section{Results}
\label{sec:results}
In this section, we present the results of our analysis on both the interaction data and the recommendation outputs using standard evaluation metrics. 
We examine how structural and fairness-related properties of the data, such as genre diversity and author gender representation, evolved over time in the data, and how these same properties are reflected in the generated recommendations. 
This analysis also allows us to assess whether any measurable changes are associated with the introduction of Goodreads’ recommender system in 2011. In each figure, the recommender introduction is marked with a triangle on the $x$-axis.

\subsection{Data}
We first explore the temporal patterns in the user interaction data to understand how its structure, diversity, and fairness properties evolved over time, regardless of the recommender outputs.
All measurements are on a monthly basis.

\begin{figure}[tb]
    \centering
    \begin{subfigure}[b]{0.45\textwidth}
        \centering
        \includegraphics[width=\textwidth]{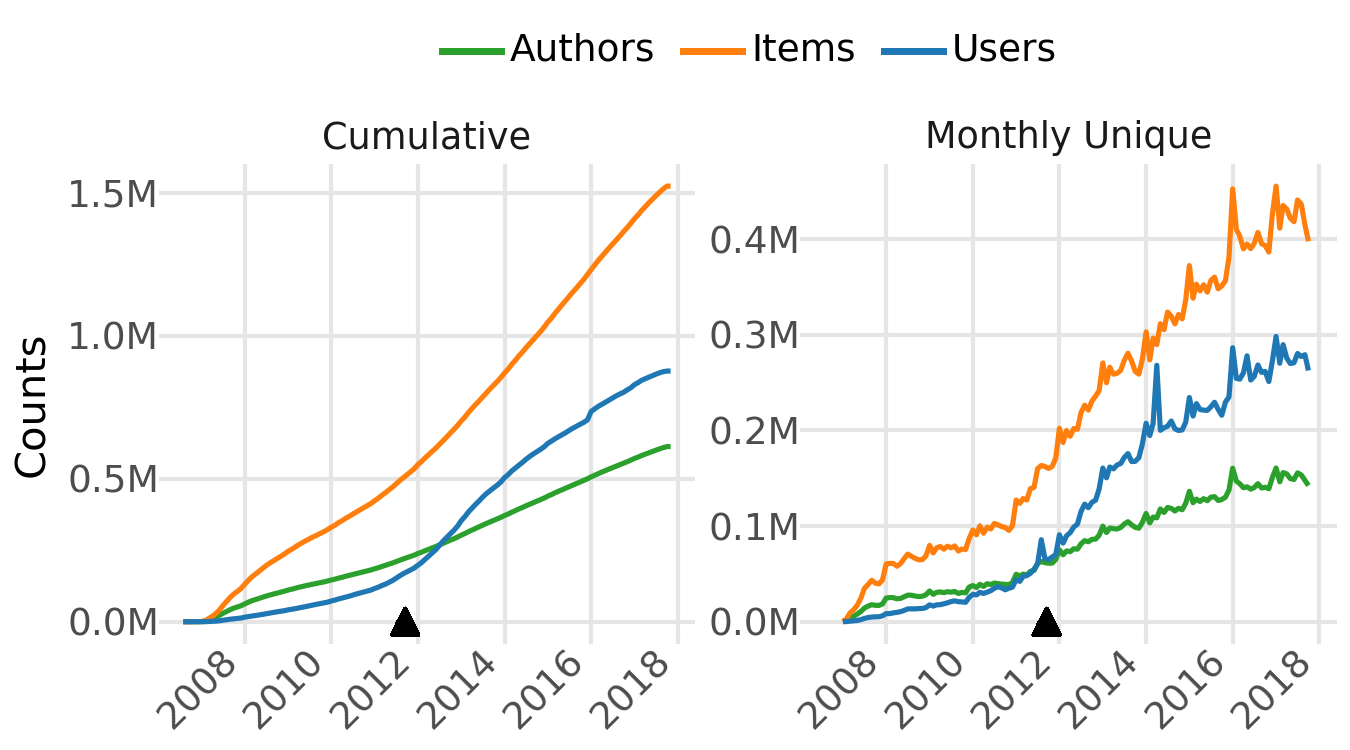}
        \caption{Unique entities}
        \label{fig:unique_count}
    \end{subfigure}
    \hfill
    \begin{subfigure}[b]{0.45\textwidth}
        \centering
        \includegraphics[width=\textwidth]{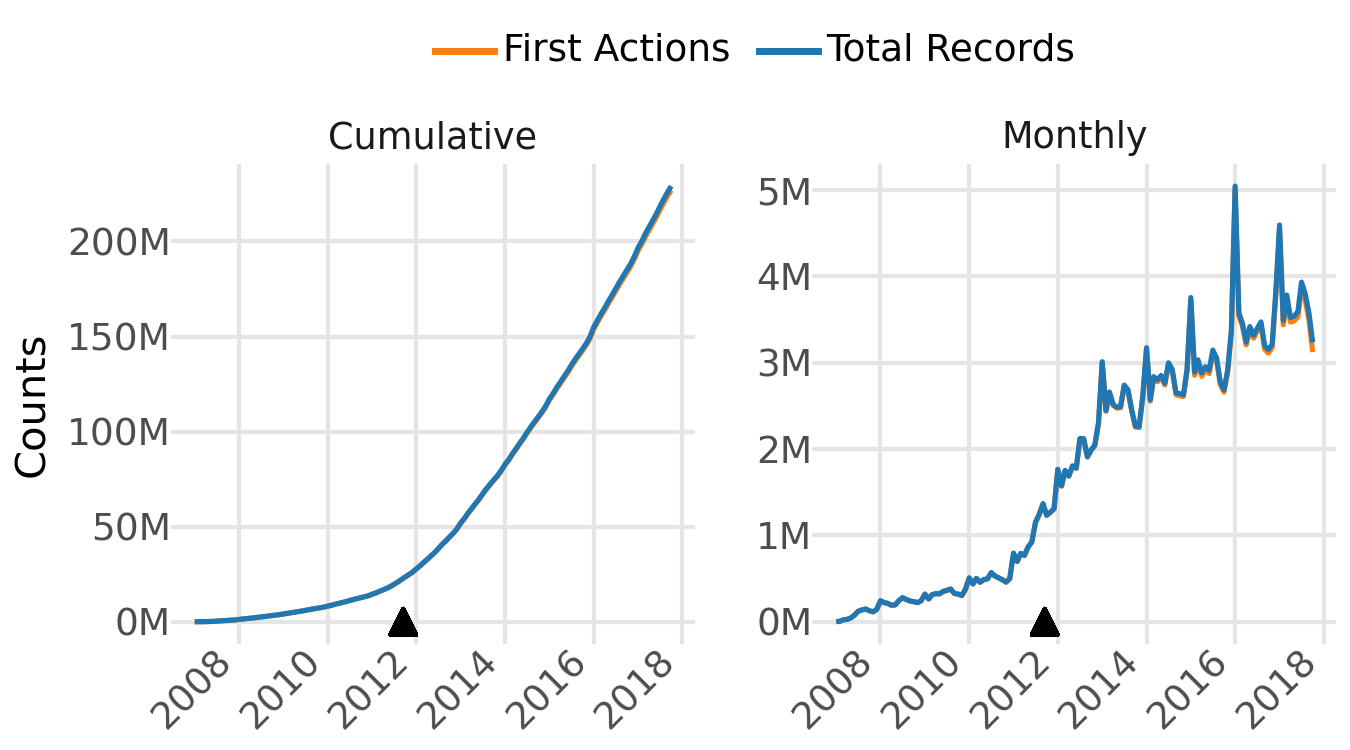}
        \caption{Interaction records}
        \label{fig:data_size}
    \end{subfigure}
    \caption{Unique entity and interaction counts per month.}
    \label{fig:combined_figure}
\end{figure}

\paragraph{Data size and growth over time} Figure~\ref{fig:combined_figure} shows the monthly activity levels in the interaction data. 
Figure~\ref{fig:unique_count} shows the monthly volume of unique users, books, and authors in the data. 
All three entities show increasing activity over time, with the sharpest increases occurring after 2011. 
Figure~\ref{fig:data_size} shows the number of records (user-book interactions) and the number of ``first'' interactions (the first time a user interacts with a particular book).  
Both lines almost completely overlap and steadily increase as expected, reflecting  the platform's growth over the ten-year period. 
While there are several fluctuations in monthly user activity, the highest peak occurs around 2016.

\paragraph{Genre diversity} To quantify genre diversity in the interaction data, we derive user-specific genre distributions by aggregating the genre distributions of the books each user has interacted with during each month. 
For a given user, the probability assigned to each genre reflects the cumulative probability mass of that genre across all books in their interaction history, based on the book-level distributions $P(g \mid b)$ defined in \S\ref{sec:methods}. 
We then compute the Shannon entropy of this user-level distribution to measure the genre diversity in their interactions, with higher entropy indicating a more diverse selection. 
The final metric is reported as the average entropy across all users. 

\begin{figure}[tb]
    {\centering
    \includegraphics[width=0.45\textwidth]{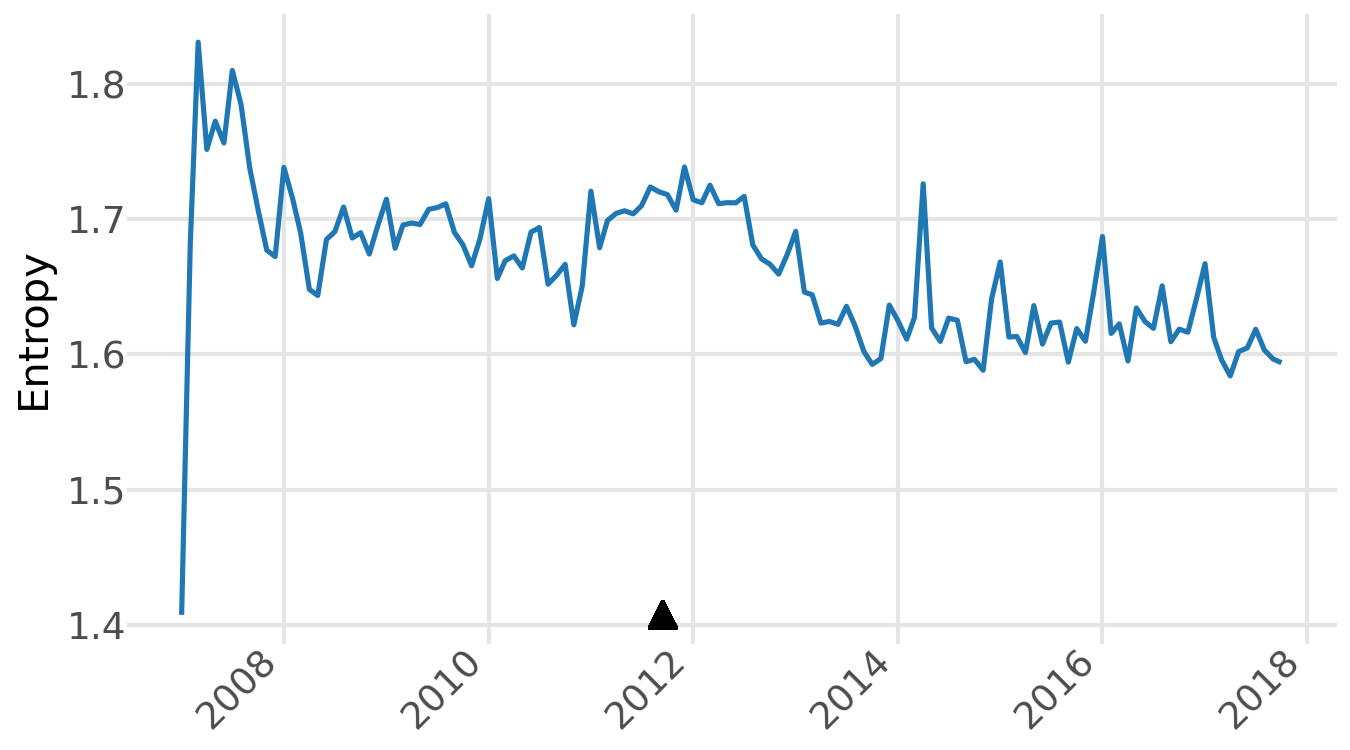}}
    \caption{Average genre entropy per user.}
    \label{fig:user-genre-entropy}
\end{figure}

Figure~\ref{fig:user-genre-entropy} shows the average genre entropy across user interactions. 
We can see that genre diversity stabilizes after an initial rise, holding relatively stable with some fluctuations until about 2012, after which it decreases. 
The patterns suggest that user interaction with genres remained relatively consistent over time.

\paragraph{Individual fairness and popularity bias} We use the \textit{Gini index} to measure the distribution of interactions across individual items and authors in the data. 
Higher Gini values indicate greater inequality, i.e., a limited number of books or authors receiving most of the interactions. 

\begin{figure}[tb]
    {\centering
    \includegraphics[width=0.45\textwidth]{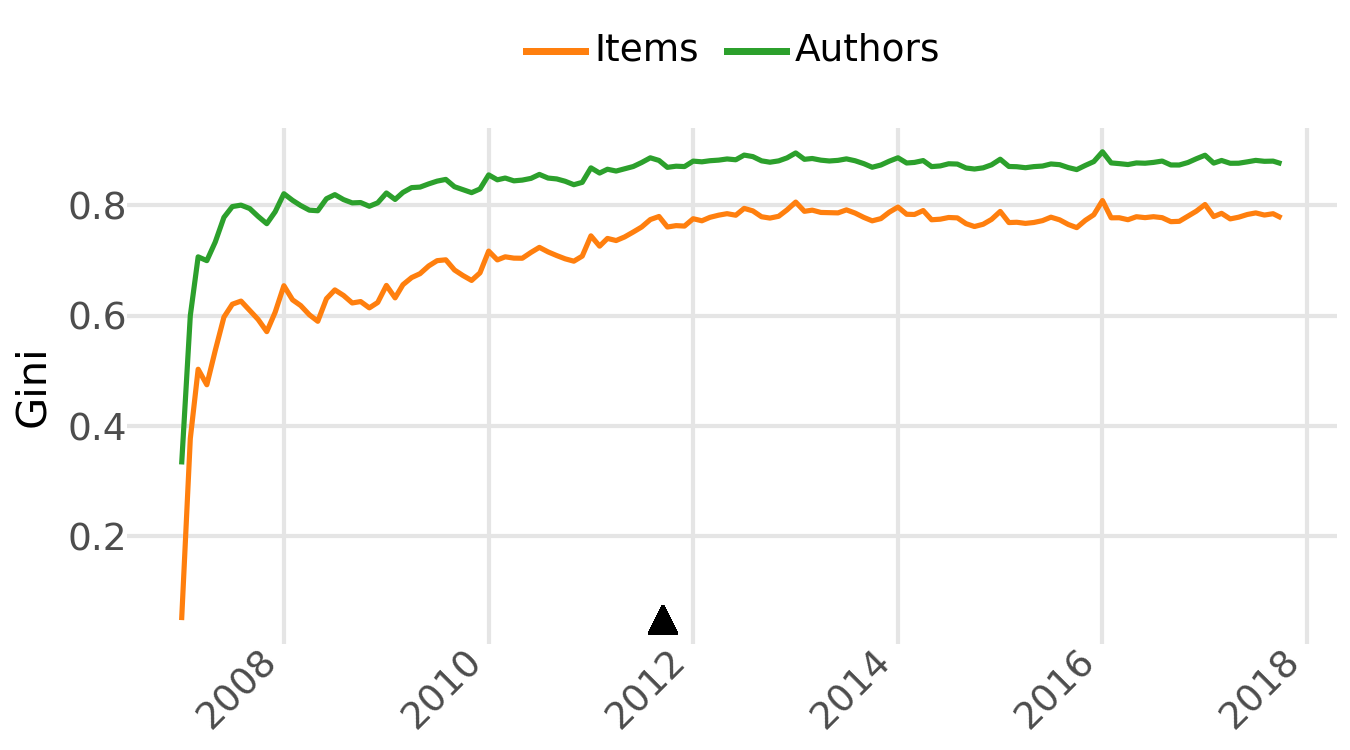}}
    \caption{Gini of monthly interactions.}
    \label{fig:gini-monthly-interactions}
\end{figure}

Figure~\ref{fig:gini-monthly-interactions} shows that book-level inequality is consistently high and relatively stable, especially after 2012, with author-level inequality even higher.
These results suggest persistent bias in user engagement toward popular books and authors.

\paragraph{Author gender representation} Finally, we examine author gender representation in user interactions by computing the monthly proportion of books interacted with that are written by female authors (only considering books for which the first author's gender identity is known).

\begin{figure}[tb]
    {\centering
    \includegraphics[width=0.45\textwidth]{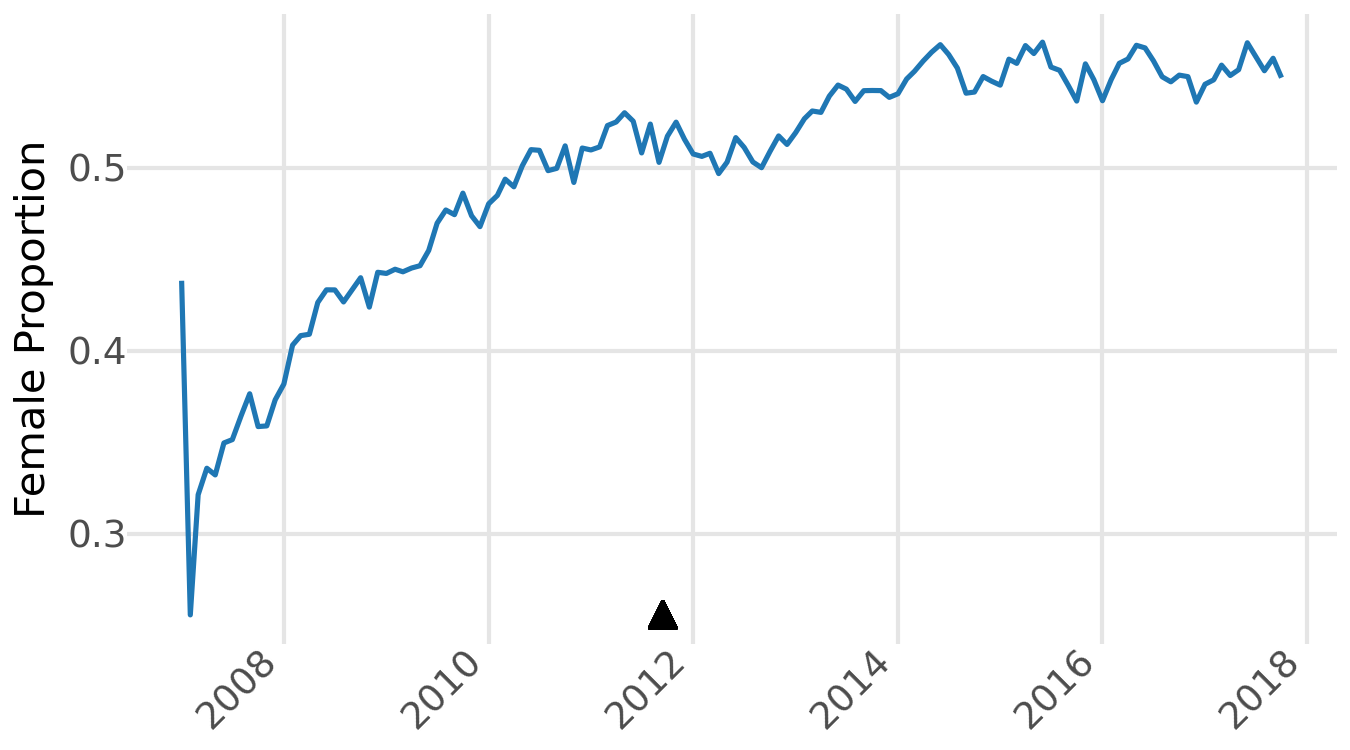}}
    \caption{Author gender representation.}
    \label{fig:author-gender}
\end{figure}

As shown in Figure~\ref{fig:author-gender}, the proportion of female authors rises sharply from around 30\% in the early years to just over 50\% by 2011, after which it remains relatively stable through 2017 with a small jump around 2014.
This is consistent with the analysis of \citet{thelwall2017goodreads} suggesting that Goodreads exhibits a female-author bias, which in turn aligns with the hypothesis that female authors were more commercially successful than male authors due to being more frequently rated online \cite{Verboord2011}, though further evidence would be needed to confirm this. 

\subsection{Recommendations}
We now turn to the evaluation of recommendations generated by our selected collaborative filters for each time window. 
All measures are computed on top-100 recommendation lists for all users in a 2-month test window. 
We evaluate these recommendation lists using the same set of metrics applied in Section 4.1, alongside the ranking performance measures and rank-biased versions of entropy and Gini. 

To account for user attention bias toward higher-ranked items in a recommendation list, we apply a position-based exposure weighting model to entropy and accuracy metrics in recommendations.
Specifically, we use the geometric cascade browsing model \citep{raj2023unified, MoffatZobel2008}, defining the exposure weight $w_i$ for position $i$ in the ranked list as:
\begin{equation}
w_i = \gamma^{(i-1)}
\label{eq:weighting-model}
\end{equation}
where \( 0 < \gamma < 1 \) is a tunable \textit{patience} parameter, which we set to LensKit's default of 0.85 in our experiments.

\begin{figure}[tb]
    \centering
    \includegraphics[width=0.45\textwidth]{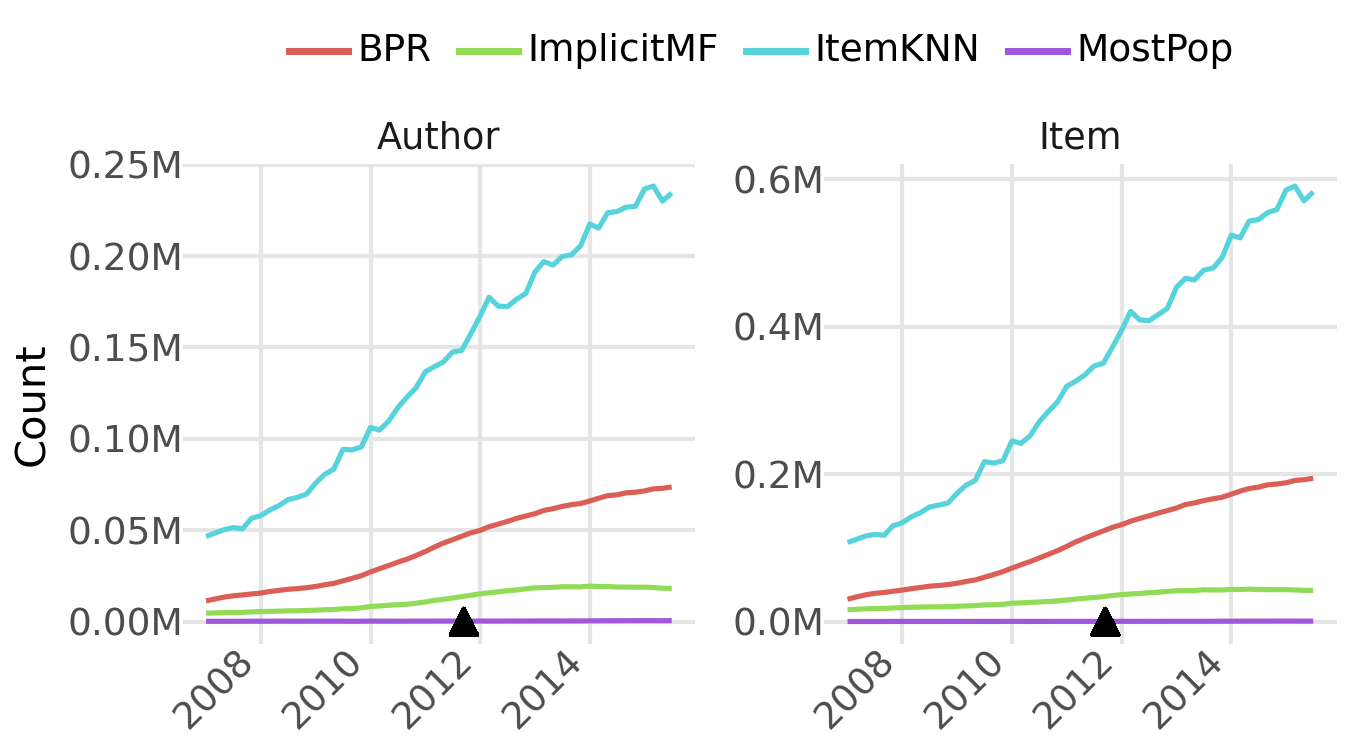}
    \caption{Monthly unique entity counts in recommendations.}
    \label{fig:unique_count_rec}
\end{figure}

\paragraph{Coverage and accuracy} Figure~\ref{fig:unique_count_rec} shows the number of unique items and authors that appear in the top-100 recommendation lists for each algorithm over time. 
ItemKNN and BPR recommend substantially more unique items and authors compared to MostPop and ImplicitMF, which is a consistent pattern over time. 
This suggests that the former algorithms provide broader coverage across both items and authors.

\begin{figure*}[tb]
    \begin{subfigure}[b]{0.33\textwidth}
    \centering\includegraphics[width=\textwidth]{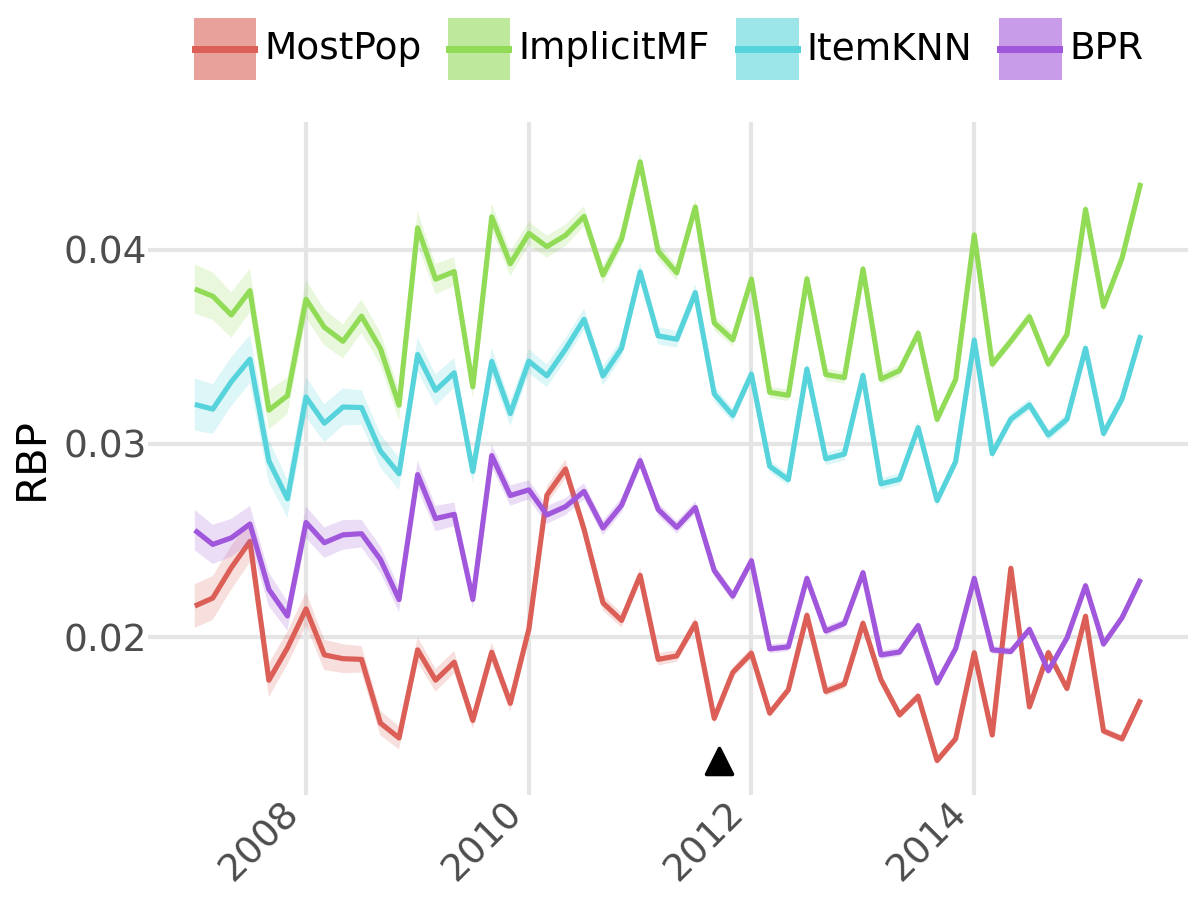}
    \caption{RBP@100}
    \label{fig:rbp}
    \end{subfigure}
    \begin{subfigure}[b]{0.33\textwidth}
    \centering\includegraphics[width=\textwidth]{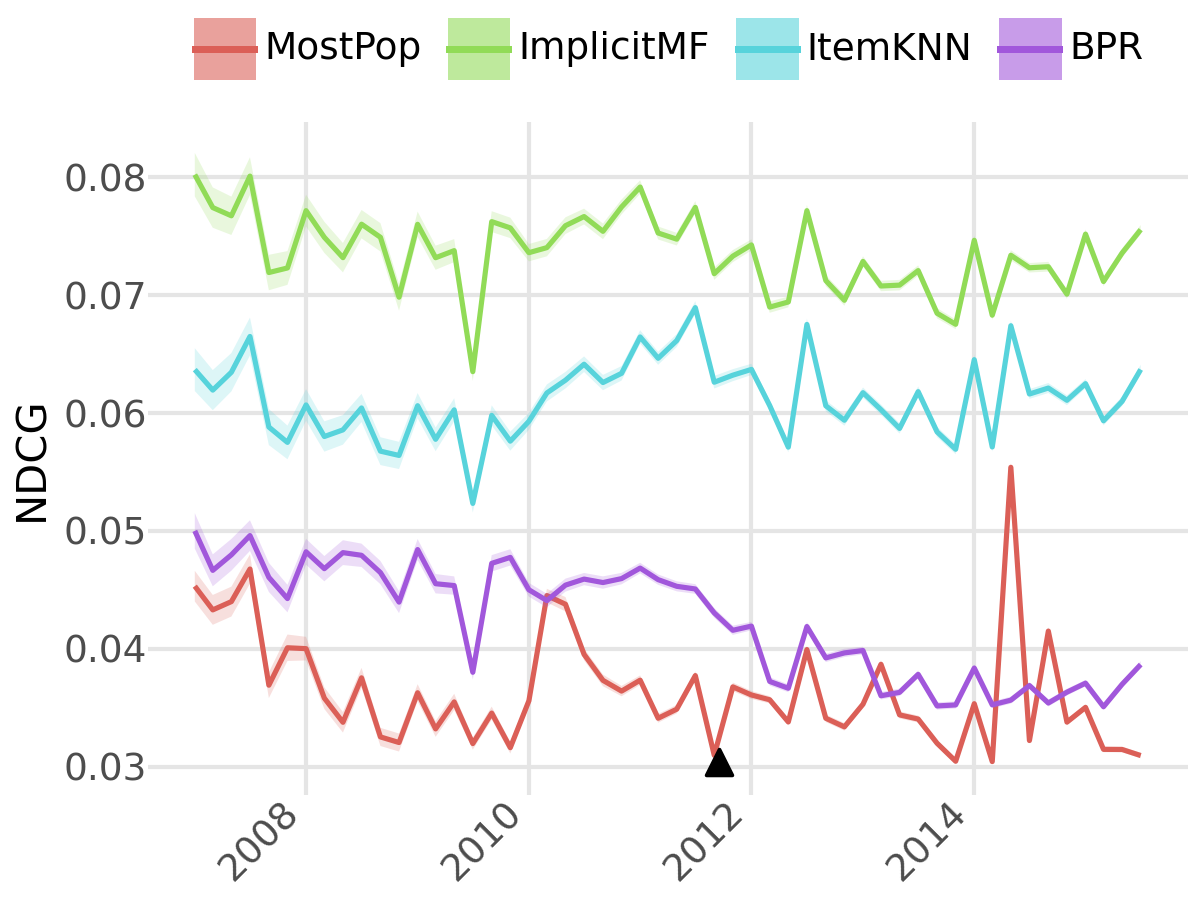}
    \caption{NDCG@100}
    \label{fig:ndcg}
    \end{subfigure}
    \begin{subfigure}[b]{0.33\textwidth}
    \centering\includegraphics[width=\textwidth]{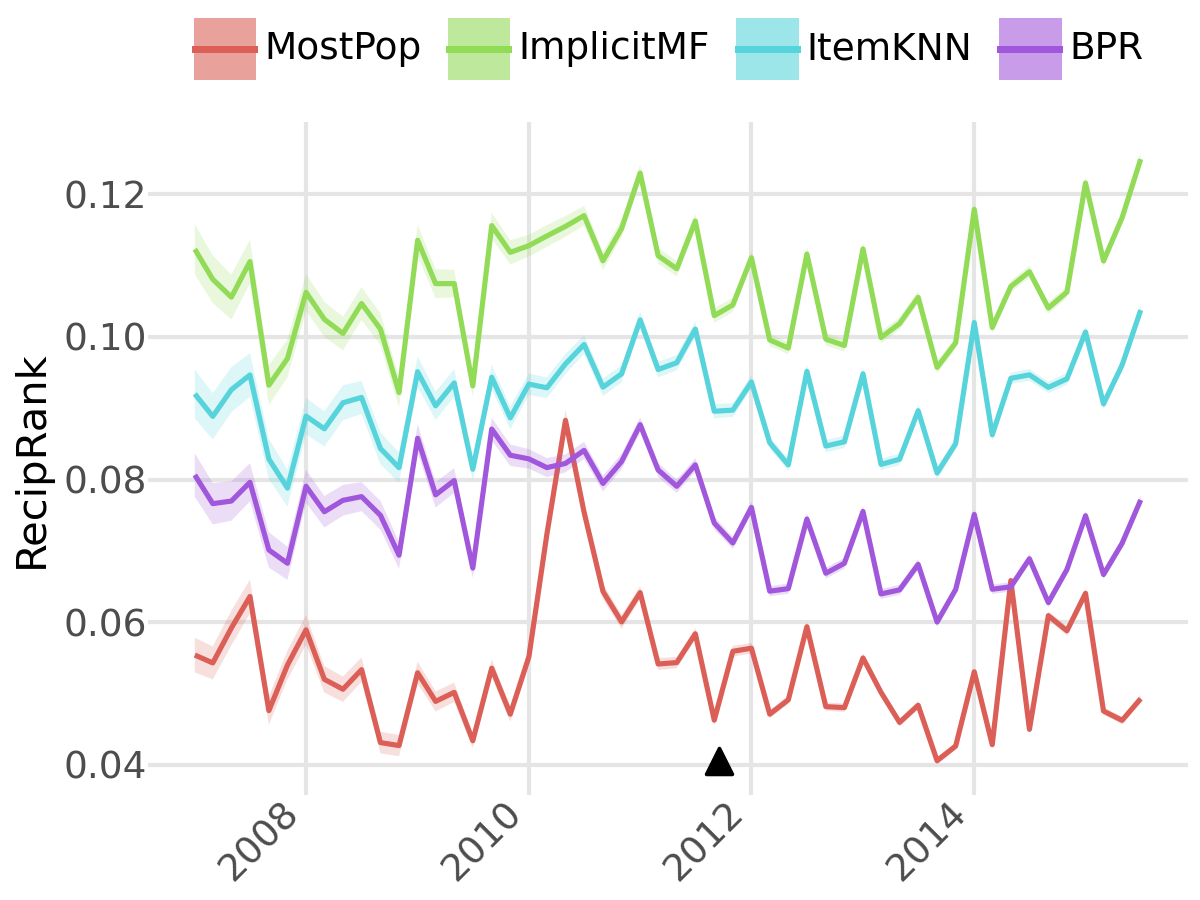}
    \caption{MRR@100}
    \label{fig:reciprank}
    \end{subfigure}
    \caption{Recommender system performance by offline metrics.}
    \label{fig:offline-perf}
\end{figure*}

We also report RBP \cite{MoffatZobel2008}, NDCG \citep{jarvelinndcg2002}, and Mean Reciprocal Rank as ranking performance metrics over time (Figures~\ref{fig:rbp},~\ref{fig:ndcg}, ~\ref{fig:reciprank}). 
ImplicitMF achieves the highest performance scores in all three metrics, followed by ItemKNN, BPR, and MostPop. 
The ordering of systems is mostly consistent over time, but there are noticeable fluctuations in performance.
The performance gap between the best-performing models (ImplicitMF and ItemKNN) and the worst-performing models (MostPop and BPR) also appears to widen over time.
The low scores of MostPop suggest that personalized models have performed better in capturing meaningful variations in user taste. 

\paragraph{Genre diversity} For each recommendation list, we compute a weighted probability distribution over genres.
The probability of genre $g$ in a user's list $L$ is defined as:

\begin{equation}
P(g \mid L) \propto \sum_{i=1}^{|L|} w_i \cdot P(g \mid L(i))
\label{eq:genre-entropy}
\end{equation}

where $w_i$ represents the exposure weight of position $i$ and $P(g \mid L(i))$ is the probability that the book in position $i$ belongs to genre $g$ (see Section 3.1). 
We then compute entropy over the resulting genre distribution to obtain the genre entropy of a list, and average across lists to obtain an overall diversity measure of \textit{mean rank-biased genre entropy}. 
As with profile entropy, larger entropy values indicate more diverse recommendation lists (i.e., more uniform distribution across genres).

\begin{figure}[tb]
    \centering
    \includegraphics[width=0.45\textwidth]{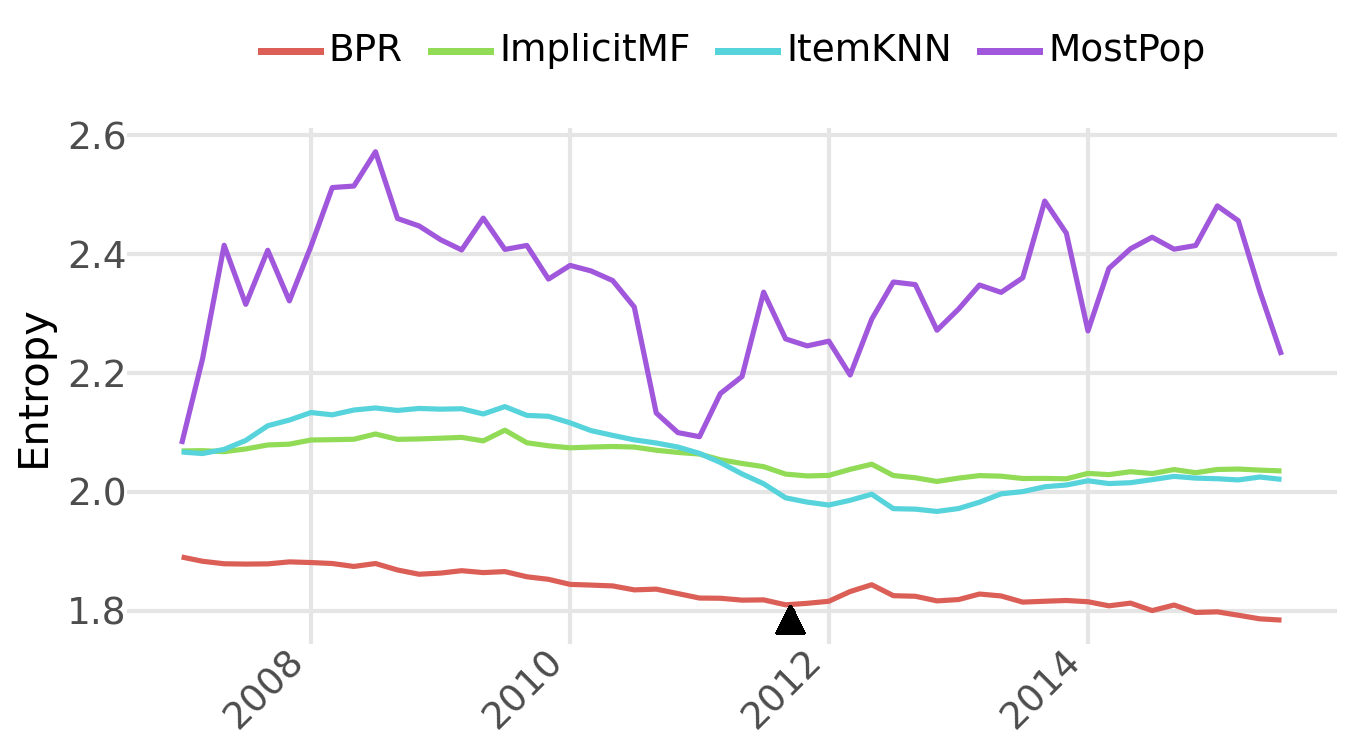}
    \caption{Genre entropy in recommendations.}
    \label{fig:rec-genre-entropy}
\end{figure}

Figure~\ref{fig:rec-genre-entropy} shows the mean genre entropy across recommendation lists.
Perhaps surprisingly, MostPop yields the highest entropy. We see at least two possible reasons for this. 
First, since MostPop recommends a fixed set of widely popular books, these items are more likely to have user-generated genre tags, possibly resulting in higher measured entropy.
In contrast, the other algorithms recommend a broader set of less popular items (as seen in Figure~\ref{fig:unique_count_rec}), which may include books with sparse or missing genre data that are excluded from the entropy calculation. 
Second, the most popular books themselves may be more diverse in genre, for example, including one highly popular book from each genre category.

BPR consistently exhibits the lowest genre diversity, likely due to its narrower focus on user-specific patterns in recommending items.
Overall, all algorithms show relatively stable or slightly declining entropy over time, with more variation in MostPop.
This suggests that personalized recommendation models are less influenced by changes in the genre distribution of the most popular books in any particular time period.

\paragraph{Individual fairness and popularity bias} We compute the Gini index (\textit{Gini@k}) over the exposure weights $w_i$, which estimate the exposure assigned to an item or author at position $i$ in the ranked recommendation lists. 
This effectively measures the allocation of ``user attention'' as a resource, accounting for the fact that users are likely to give more of their attention to books at the top of a recommendation list.

\begin{figure}[tb]
    \centering    \includegraphics[width=0.45\textwidth]{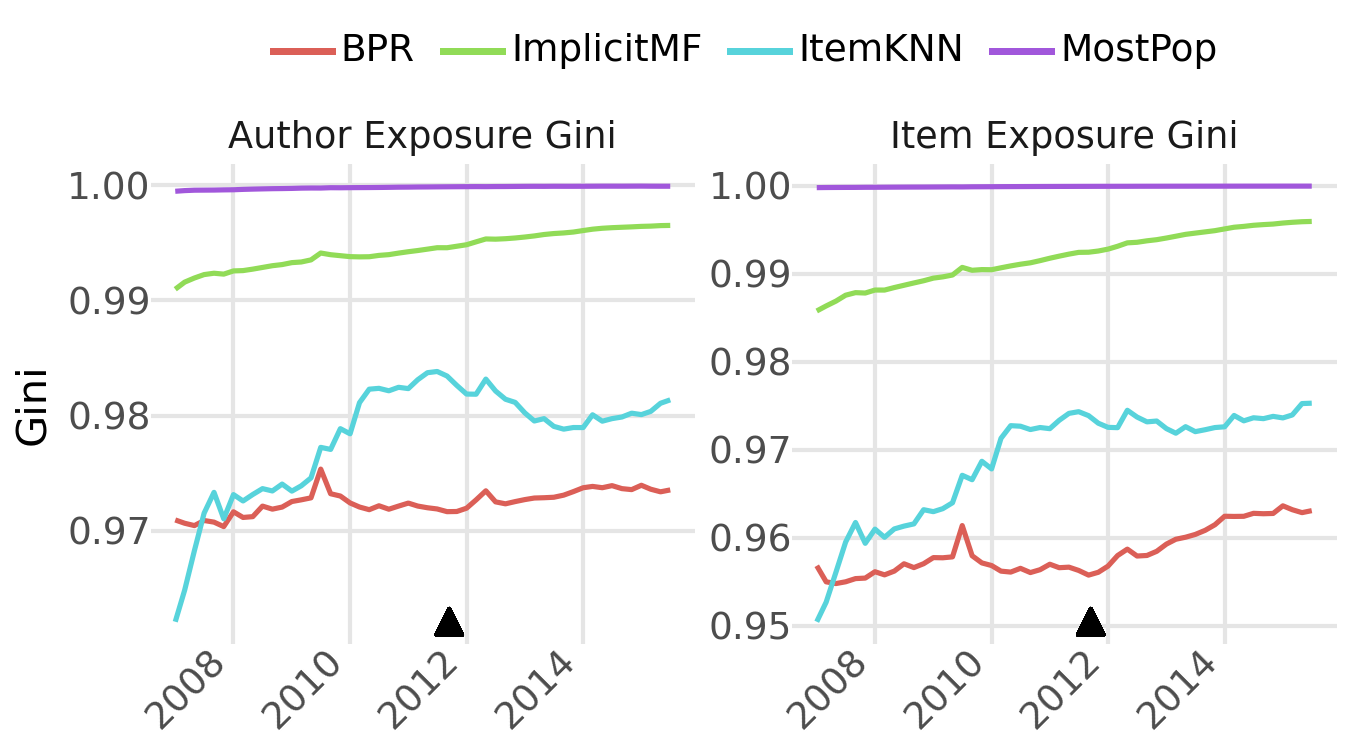}
    \caption{Item and author popularity concentration (Gini @ 100).}
    \label{fig:gini}
\end{figure}

Figure~\ref{fig:gini} shows the Gini index for author-level (left) and item-level (right) exposure. 
As expected, MostPop displays the highest inequality in exposure, with both item- and author-level Gini values consistently close to 1.0.
ItemKNN and BPR produce more balanced exposure, especially in earlier years.
The ordering of ItemKNN and ImplicitMF we observe here differs from that found in the movie recommendation by \citet{ekstranddistributionalrs2024}, indicating that models' relative fairness properties may differ between domains or datasets.

All models show a gradual increase in Gini values over time and maintain relatively high levels of exposure inequality. 
This indicates a growing exposure bias toward a smaller set of items and authors, even in more personalized algorithms.
The high Gini values are also influenced by the domain of items: for consistent comparison across both models and time windows, we computed Gini over all items in the dataset, with the unrecommended items receiving zero exposure.

\begin{figure}[ht]
    \centering
    \includegraphics[width=0.45\textwidth]{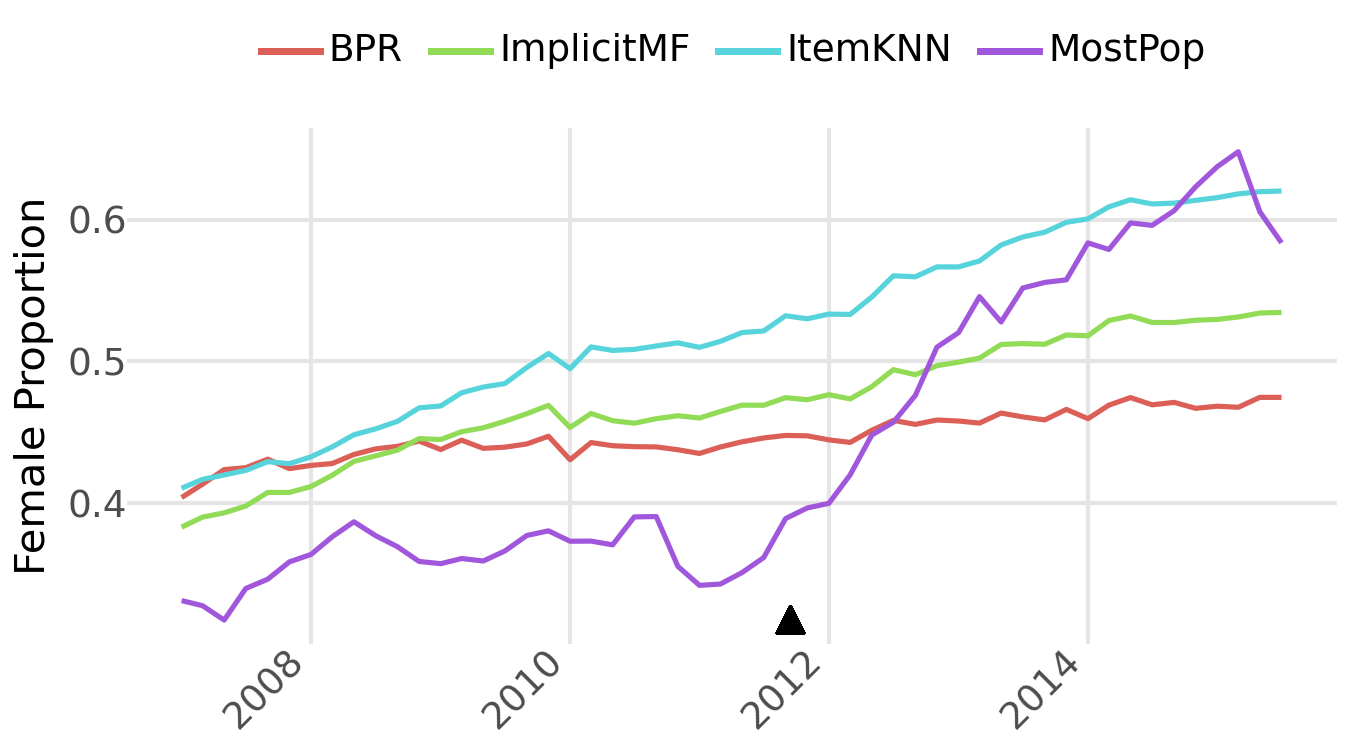}
    \caption{Author gender representation in recommendations.}
    \label{fig:rec-author-gender}
\end{figure}

\paragraph{Author gender representation} 
Figure~\ref{fig:rec-author-gender} shows the proportion of books in the recommendations that have female authors. 
ItemKNN shows the highest and most steady increase in female author representation over time. 
ImplicitMF also increases gradually, while BPR stays more stable with a smaller rise. 
MostPop starts with the lowest proportion and increases slowly until a sharp jump near the end. 
All models show upward trends, but the speed and pattern of growth are different. 
The overall trend reflects the underlying patterns seen in the interaction data discussed in Section 4.1.

\subsection{Effects of Goodreads Recommendation}
To evaluate whether the introduction of Goodreads’ recommender system in 2011 is associated with measurable changes in system behavior, we analyzed temporal trends in both the interaction data and the outputs of our trained models.
Overall, we did not observe clear or immediate changes in diversity, fairness, or effectiveness metrics around 2011 that can be directly associated with the launch of the recommender system.

However, there are some patterns that are worth further investigation. 
One such pattern is a gradual decrease in genre entropy in the interaction data after 2012, which may suggest an increasing influence of recommendation or narrowing user exploration. 
While the models differ in how they reflect underlying data trends, ImplicitMF and ItemKNN show signs of stabilization in both genre entropy and exposure Gini metrics following the system’s introduction. 
We also did not see identifiable changes in recommender effectiveness associated with the introduction of algorithmic recommendation; increased effectiveness on offline evaluation metrics may be a sign of user interaction behavior becoming more predictable (due to the influence of algorithmic recommendations), but we do not see clear evidence of such an effect.

Overall, the observed trends appear to reflect broader, ongoing developments in the system, such as the accumulation of user interaction history, growth of the item catalog, and evolving user behavior, rather than the effects of a single intervention.
Existing-system bias (bias in item exposure due to the recommender system being active while data is collected) is a significant problem in offline recommender system evaluation \citep{algorithmicconfoundingchaney2018, chenBiasDebiasRecommender2023, jeunenRevisitingOfflineEvaluation2019}.
Goodreads' lack of algorithmic recommendations for its first 4--5 years of operation provides an opportunity to detect potential existing-system biases.
Our analyses were not able to detect specific problems attributable to the addition of a recommender system in this particular dataset.
Algorithmic recommendations have significantly less prominence in the Goodreads user experience than they do in many other platforms, so they may have less of an impact on user behavior.

\section{Discussion}
\label{sec:discussion}
We observed different trends in diversity and fairness measures across both the interaction data and the generated recommendations.
We were particularly interested in understanding how aspects of the data that affect bias, fairness, diversity, and recommender performance changed over the history of the platform.

While user activity and the item corpus have increased over time, the popularity concentration in the interaction data (as measured by the Gini index) remains relatively stable, especially after 2012. 
This suggests that increased user engagement is not solely concentrated on the most popular books, but that users have had a relatively steady spread of interest in both more- and less-popular books over time.
Moreover, although the interaction Gini stabilized after 2012, the exposure Gini for models like BPR and ImplicitMF continues to rise. 
This raises the question of whether there are other patterns in user interactions that lead to increased concentration in model outputs despite stable patterns in the underlying data. Future research should investigate this possibility to better understand the drivers of popularity bias in collaborative filtering outputs.

For genre diversity, the models largely mirror the gradual changes seen in the interaction data. 
However, there is a slight difference around 2012: 
while genre entropy in user interactions peaks around that time, most algorithms show a decline. 
One possible explanation is that changes in user behavior took time to influence model training data, leading to a delayed rise in entropy in the recommendations. 
Most of the models begin showing upward trends after this point, which may reflect a lagged response to shifts in genre diversity in the data.

Our findings, in general, and particularly the changing patterns in recommender system effectiveness metrics, indicate that evaluation outcomes are dependent on the choice of test and train time windows.
This means the selection of time windows is important for the comparative evaluation of recommender systems, and suggests that results may be time-specific and potentially difficult to generalize.
While we did see a few inversions of recommendation model performance rankings, we do observe a shift in the performance gap between models as time progresses in the dataset.
This will influence the effect sizes reported in experiments on different temporal test windows.
Temporal, data-centered audits of recommender systems can uncover evolving patterns in performance, fairness, and diversity that may not be visible in static, snapshot evaluations. 

Our analysis does have several limitations.
Some of our results are affected by sparse or incomplete metadata, such as missing genre tags or author information, which limit the accuracy of certain measures.
The UCSD book graph also only contains public profiles that were still active in 2017 or 2018, so data from private or deleted accounts is not available.
We also do not know what specific recommendation techniques Goodreads deployed in 2011 (or later), so we cannot model the performance changes in their recommendation models or directly test for their influence on the data.
Finally, our results are observational and correlational, so we cannot establish causality between any of our metrics or between the introduction of the Goodreads recommender system and any user behavior changes.
Nevertheless, we believe such descriptive and exploratory analyses are valuable for understanding the contours of the data used by the recommender systems and user modeling research communities to train and evaluate models and to study the behavior and social impacts of humans and algorithms on social data.
Our findings also point to the need for further exploration of underlying drivers such as user engagement, metadata quality, and patterns our metrics could not detect but which nonetheless influence recommendation models over time.

\section{Conclusion}
\label{sec:conclusion}
Documenting and understanding datasets is important for designing experiments and contextualizing analyses and findings.
The evolution of recommender systems datasets over time is often overlooked, but it is crucial for informing time-based experimental design decisions and provides context both for interpreting experimental results and for understanding historical changes when performing simulation studies of potential future dynamics \citep{ferraro2021breaktheloop, ferraro2024itsnotyou, algorithmicconfoundingchaney2018}.

We provided such an analysis for the UCSD Book Graph, documenting both the evolution of the dataset itself and the behavior of collaborative filtering models retrained at different time points.
We found increasing representation of female authors, decreasing genre diversity, and slightly increasing concentration of interactions on popular users and items over the history of the dataset.
We did not observe any clear impacts of the introduction of algorithmic recommendations on the behaviors of either users or collaborative filtering recommender models.

Our analysis lays the retrospective groundwork for future work on the temporal dynamics of recommender system behavior, including both traditional accuracy concerns and social impact considerations such as diversity and fairness.
They also highlight the potential impact of the choice of splitting time on recommender experimental results.
We encourage groups preparing and releasing new datasets for recommender systems research to include temporal analyses along with the descriptive statistics they publish with their datasets.

\begin{acks}
Compute time was provided by the Dragon Deep Learning Cluster, funded by NSF grant 23-20600.
\end{acks}

\bibliographystyle{ACM-Reference-Format}
\bibliography{goodreadsfairumap}


\begin{thebibliography}{43}


\ifx \showCODEN    \undefined \def \showCODEN     #1{\unskip}     \fi
\ifx \showISBNx    \undefined \def \showISBNx     #1{\unskip}     \fi
\ifx \showISBNxiii \undefined \def \showISBNxiii  #1{\unskip}     \fi
\ifx \showISSN     \undefined \def \showISSN      #1{\unskip}     \fi
\ifx \showLCCN     \undefined \def \showLCCN      #1{\unskip}     \fi
\ifx \shownote     \undefined \def \shownote      #1{#1}          \fi
\ifx \showarticletitle \undefined \def \showarticletitle #1{#1}   \fi
\ifx \showURL      \undefined \def \showURL       {\relax}        \fi
\providecommand\bibfield[2]{#2}
\providecommand\bibinfo[2]{#2}
\providecommand\natexlab[1]{#1}
\providecommand\showeprint[2][]{arXiv:#2}

\bibitem[goo(2011)]%
        {goodreads2011}
 \bibinfo{year}{2011}\natexlab{}.
\newblock \bibinfo{title}{Announcing Goodreads Personalized Recommendations}.
\newblock
\newblock
\shownote{https://www.goodreads.com/blog/show/303-announcing-goodreads-personalized-recommendations}.


\bibitem[Abdollahpouri et~al\mbox{.}(2017)]%
        {abdollahpouri2017popularitybias}
\bibfield{author}{\bibinfo{person}{Himan Abdollahpouri}, \bibinfo{person}{Robin Burke}, {and} \bibinfo{person}{Bamshad Mobasher}.} \bibinfo{year}{2017}\natexlab{}.
\newblock \showarticletitle{Controlling Popularity Bias in Learning-to-Rank Recommendation}. In \bibinfo{booktitle}{\emph{Proceedings of the Eleventh ACM Conference on Recommender Systems}} \emph{(\bibinfo{series}{RecSys '17})}. \bibinfo{pages}{42–46}.
\newblock
\showISBNx{9781450346528}
\href{https://doi.org/10.1145/3109859.3109912}{doi:\nolinkurl{10.1145/3109859.3109912}}


\bibitem[Beutel et~al\mbox{.}(2019)]%
        {beutel2019fairness}
\bibfield{author}{\bibinfo{person}{Alex Beutel}, \bibinfo{person}{Jilin Chen}, \bibinfo{person}{Tulsee Doshi}, \bibinfo{person}{Hai Qian}, \bibinfo{person}{Li Wei}, \bibinfo{person}{Yi Wu}, \bibinfo{person}{Lukasz Heldt}, \bibinfo{person}{Zhe Zhao}, \bibinfo{person}{Lichan Hong}, \bibinfo{person}{Ed~H Chi}, {et~al\mbox{.}}} \bibinfo{year}{2019}\natexlab{}.
\newblock \showarticletitle{Fairness in Recommendation Ranking Through Pairwise Comparisons}. In \bibinfo{booktitle}{\emph{Proceedings of the 25th ACM SIGKDD International Conference on Knowledge Discovery \& Data Mining}}. \bibinfo{pages}{2212--2220}.
\newblock
\href{https://doi.org/10.1145/3292500.3330745}{doi:\nolinkurl{10.1145/3292500.3330745}}


\bibitem[Boka et~al\mbox{.}(2024)]%
        {boka2024survey}
\bibfield{author}{\bibinfo{person}{Tesfaye~Fenta Boka}, \bibinfo{person}{Zhendong Niu}, {and} \bibinfo{person}{Rama~Bastola Neupane}.} \bibinfo{year}{2024}\natexlab{}.
\newblock \showarticletitle{A Survey of Sequential Recommendation Systems: Techniques, Evaluation, and Future Direction}.
\newblock \bibinfo{journal}{\emph{Information Systems}} (\bibinfo{year}{2024}), \bibinfo{pages}{102427}.
\newblock
\href{https://doi.org/10.1016/j.is.2024.102427}{doi:\nolinkurl{10.1016/j.is.2024.102427}}


\bibitem[Burke(2010)]%
        {burkeEvaluatingDynamicProperties2010}
\bibfield{author}{\bibinfo{person}{Robin Burke}.} \bibinfo{year}{2010}\natexlab{}.
\newblock \showarticletitle{Evaluating the Dynamic Properties of Recommendation Algorithms}. In \bibinfo{booktitle}{\emph{Proceedings of the {{Fourth ACM Conference}} on {{Recommender Systems}}}}. \bibinfo{pages}{225--228}.
\newblock
\href{https://doi.org/10.1145/1864708.1864753}{doi:\nolinkurl{10.1145/1864708.1864753}}


\bibitem[Chaney et~al\mbox{.}(2018)]%
        {algorithmicconfoundingchaney2018}
\bibfield{author}{\bibinfo{person}{Allison J.~B. Chaney}, \bibinfo{person}{Brandon~M. Stewart}, {and} \bibinfo{person}{Barbara~E. Engelhardt}.} \bibinfo{year}{2018}\natexlab{}.
\newblock \showarticletitle{How Algorithmic Confounding in Recommendation Systems Increases Homogeneity and Decreases Utility}. In \bibinfo{booktitle}{\emph{Proceedings of the 12th ACM Conference on Recommender Systems}} \emph{(\bibinfo{series}{RecSys '18})}. \bibinfo{pages}{224–232}.
\newblock
\showISBNx{9781450359016}
\href{https://doi.org/10.1145/3240323.3240370}{doi:\nolinkurl{10.1145/3240323.3240370}}


\bibitem[Chen et~al\mbox{.}(2023)]%
        {chenBiasDebiasRecommender2023}
\bibfield{author}{\bibinfo{person}{Jiawei Chen}, \bibinfo{person}{Hande Dong}, \bibinfo{person}{Xiang Wang}, \bibinfo{person}{Fuli Feng}, \bibinfo{person}{Meng Wang}, {and} \bibinfo{person}{Xiangnan He}.} \bibinfo{year}{2023}\natexlab{}.
\newblock \showarticletitle{Bias and Debias in Recommender System: A Survey and Future Directions}.
\newblock \bibinfo{journal}{\emph{ACM Transactions on Information Systems}} \bibinfo{volume}{41}, \bibinfo{number}{3} (\bibinfo{date}{Feb.} \bibinfo{year}{2023}), \bibinfo{pages}{1--39}.
\newblock
\showISSN{1094-9224}
\href{https://doi.org/10.1145/3564284}{doi:\nolinkurl{10.1145/3564284}}


\bibitem[Deldjoo et~al\mbox{.}(2024)]%
        {deldjoo2024fairness}
\bibfield{author}{\bibinfo{person}{Yashar Deldjoo}, \bibinfo{person}{Dietmar Jannach}, \bibinfo{person}{Alejandro Bellogin}, \bibinfo{person}{Alessandro Difonzo}, {and} \bibinfo{person}{Dario Zanzonelli}.} \bibinfo{year}{2024}\natexlab{}.
\newblock \showarticletitle{Fairness in Recommender Systems: Research Landscape and Future Directions}.
\newblock \bibinfo{journal}{\emph{User Modeling and User-Adapted Interaction}} \bibinfo{volume}{34}, \bibinfo{number}{1} (\bibinfo{year}{2024}), \bibinfo{pages}{59--108}.
\newblock
\href{https://doi.org/10.1007/s11257-023-09364-z}{doi:\nolinkurl{10.1007/s11257-023-09364-z}}


\bibitem[Deshpande and Karypis(2004)]%
        {deshpandeItembasedTopRecommendation2004}
\bibfield{author}{\bibinfo{person}{Mukund Deshpande} {and} \bibinfo{person}{George Karypis}.} \bibinfo{year}{2004}\natexlab{}.
\newblock \showarticletitle{Item-Based Top-N Recommendation Algorithms}.
\newblock \bibinfo{journal}{\emph{ACM Transactions on Information Systems}} \bibinfo{volume}{22}, \bibinfo{number}{1} (\bibinfo{year}{2004}), \bibinfo{pages}{143--177}.
\newblock
\showISSN{1046-8188, 1558-2868}
\href{https://doi.org/10.1145/963770.963776}{doi:\nolinkurl{10.1145/963770.963776}}


\bibitem[Ekstrand(2020)]%
        {ekstrandLensKitPythonNextGeneration2020}
\bibfield{author}{\bibinfo{person}{Michael~D. Ekstrand}.} \bibinfo{year}{2020}\natexlab{}.
\newblock \showarticletitle{{{LensKit}} for {{Python}}: {{Next-Generation Software}} for {{Recommender Systems Experiments}}}. In \bibinfo{booktitle}{\emph{Proceedings of the 29th {{ACM International Conference}} on {{Information}} \& {{Knowledge Management}}}} (2020-10-19). \bibinfo{pages}{2999--3006}.
\newblock
\showISBNx{978-1-4503-6859-9}
\href{https://doi.org/10.1145/3340531.3412778}{doi:\nolinkurl{10.1145/3340531.3412778}}


\bibitem[Ekstrand et~al\mbox{.}(2024)]%
        {ekstranddistributionalrs2024}
\bibfield{author}{\bibinfo{person}{Michael~D. Ekstrand}, \bibinfo{person}{Ben Carterette}, {and} \bibinfo{person}{Fernando Diaz}.} \bibinfo{year}{2024}\natexlab{}.
\newblock \showarticletitle{Distributionally-Informed Recommender System Evaluation}.
\newblock \bibinfo{journal}{\emph{ACM Transactions on Recommender Systems}} \bibinfo{volume}{2}, \bibinfo{number}{1}, Article \bibinfo{articleno}{6} (\bibinfo{year}{2024}).
\newblock
\href{https://doi.org/10.1145/3613455}{doi:\nolinkurl{10.1145/3613455}}


\bibitem[Ekstrand et~al\mbox{.}(2022)]%
        {ekstrand2022fairness}
\bibfield{author}{\bibinfo{person}{Michael~D Ekstrand}, \bibinfo{person}{Anubrata Das}, \bibinfo{person}{Robin Burke}, \bibinfo{person}{Fernando Diaz}, {et~al\mbox{.}}} \bibinfo{year}{2022}\natexlab{}.
\newblock \showarticletitle{Fairness in Information Access Systems}.
\newblock \bibinfo{journal}{\emph{Foundations and Trends{\textregistered} in Information Retrieval}} \bibinfo{volume}{16}, \bibinfo{number}{1-2} (\bibinfo{year}{2022}), \bibinfo{pages}{1--177}.
\newblock
\href{https://doi.org/10.1561/1500000079}{doi:\nolinkurl{10.1561/1500000079}}


\bibitem[Ekstrand and Kluver(2021)]%
        {ekstrand2021exploring}
\bibfield{author}{\bibinfo{person}{Michael~D. Ekstrand} {and} \bibinfo{person}{Daniel Kluver}.} \bibinfo{year}{2021}\natexlab{}.
\newblock \showarticletitle{Exploring Author Gender in Book Rating and Recommendation}.
\newblock \bibinfo{journal}{\emph{User Modeling and User-Adapted Interaction}}  \bibinfo{volume}{31} (\bibinfo{year}{2021}), \bibinfo{pages}{377--420}.
\newblock
\href{https://doi.org/10.1007/s11257-020-09284-2}{doi:\nolinkurl{10.1007/s11257-020-09284-2}}


\bibitem[Fabbri et~al\mbox{.}(2022)]%
        {fabbri2022exposure}
\bibfield{author}{\bibinfo{person}{Francesco Fabbri}, \bibinfo{person}{Maria~Luisa Croci}, \bibinfo{person}{Francesco Bonchi}, {and} \bibinfo{person}{Carlos Castillo}.} \bibinfo{year}{2022}\natexlab{}.
\newblock \showarticletitle{Exposure Inequality in People Recommender Systems: The Long-Term Effects}. In \bibinfo{booktitle}{\emph{Proceedings of the International AAAI Conference on Web and Social Media}}, Vol.~\bibinfo{volume}{16}. \bibinfo{pages}{194--204}.
\newblock
\href{https://doi.org/10.1609/icwsm.v16i1.19284}{doi:\nolinkurl{10.1609/icwsm.v16i1.19284}}


\bibitem[Ferraro et~al\mbox{.}(2024)]%
        {ferraro2024itsnotyou}
\bibfield{author}{\bibinfo{person}{Andres Ferraro}, \bibinfo{person}{Michael~D. Ekstrand}, {and} \bibinfo{person}{Christine Bauer}.} \bibinfo{year}{2024}\natexlab{}.
\newblock \showarticletitle{It's Not You, It's Me: The Impact of Choice Models and Ranking Strategies on Gender Imbalance in Music Recommendation}. In \bibinfo{booktitle}{\emph{Proceedings of the 18th ACM Conference on Recommender Systems}} \emph{(\bibinfo{series}{RecSys '24})}. \bibinfo{pages}{884–889}.
\newblock
\showISBNx{9798400705052}
\href{https://doi.org/10.1145/3640457.3688163}{doi:\nolinkurl{10.1145/3640457.3688163}}


\bibitem[Ferraro et~al\mbox{.}(2021)]%
        {ferraro2021breaktheloop}
\bibfield{author}{\bibinfo{person}{Andres Ferraro}, \bibinfo{person}{Xavier Serra}, {and} \bibinfo{person}{Christine Bauer}.} \bibinfo{year}{2021}\natexlab{}.
\newblock \showarticletitle{Break the Loop: Gender Imbalance in Music Recommenders}. In \bibinfo{booktitle}{\emph{Proceedings of the 2021 Conference on Human Information Interaction and Retrieval}} \emph{(\bibinfo{series}{CHIIR '21})}. \bibinfo{pages}{249–254}.
\newblock
\showISBNx{9781450380553}
\href{https://doi.org/10.1145/3406522.3446033}{doi:\nolinkurl{10.1145/3406522.3446033}}


\bibitem[Ge et~al\mbox{.}(2021)]%
        {ge2021towards}
\bibfield{author}{\bibinfo{person}{Yingqiang Ge}, \bibinfo{person}{Shuchang Liu}, \bibinfo{person}{Ruoyuan Gao}, \bibinfo{person}{Yikun Xian}, \bibinfo{person}{Yunqi Li}, \bibinfo{person}{Xiangyu Zhao}, \bibinfo{person}{Changhua Pei}, \bibinfo{person}{Fei Sun}, \bibinfo{person}{Junfeng Ge}, \bibinfo{person}{Wenwu Ou}, {et~al\mbox{.}}} \bibinfo{year}{2021}\natexlab{}.
\newblock \showarticletitle{Towards Long-Term Fairness in Recommendation}. In \bibinfo{booktitle}{\emph{Proceedings of the 14th ACM International Conference on Web Search and Data Mining}}. \bibinfo{pages}{445--453}.
\newblock
\href{https://doi.org/10.1145/3437963.344182}{doi:\nolinkurl{10.1145/3437963.344182}}


\bibitem[Gebru et~al\mbox{.}(2021)]%
        {gebru2021datasheets}
\bibfield{author}{\bibinfo{person}{Timnit Gebru}, \bibinfo{person}{Jamie Morgenstern}, \bibinfo{person}{Briana Vecchione}, \bibinfo{person}{Jennifer~Wortman Vaughan}, \bibinfo{person}{Hanna Wallach}, \bibinfo{person}{Hal~Daum{\'e} Iii}, {and} \bibinfo{person}{Kate Crawford}.} \bibinfo{year}{2021}\natexlab{}.
\newblock \showarticletitle{Datasheets for Datasets}.
\newblock \bibinfo{journal}{\emph{Commun. ACM}} \bibinfo{volume}{64}, \bibinfo{number}{12} (\bibinfo{year}{2021}), \bibinfo{pages}{86--92}.
\newblock
\href{https://doi.org/10.1145/345872}{doi:\nolinkurl{10.1145/345872}}


\bibitem[Giner-Miguelez et~al\mbox{.}(2022)]%
        {giner2022describeml}
\bibfield{author}{\bibinfo{person}{Joan Giner-Miguelez}, \bibinfo{person}{Abel G{\'o}mez}, {and} \bibinfo{person}{Jordi Cabot}.} \bibinfo{year}{2022}\natexlab{}.
\newblock \showarticletitle{DescribeML: A Tool for Describing Machine Learning Dataset}. In \bibinfo{booktitle}{\emph{Proceedings of the 25th International Conference on Model Driven Engineering Languages and Systems: Companion Proceedings}}. \bibinfo{pages}{22--26}.
\newblock
\href{https://doi.org/10.1145/3550356.3559087}{doi:\nolinkurl{10.1145/3550356.3559087}}


\bibitem[Gundersen and Kjensmo(2018)]%
        {gundersen2018state}
\bibfield{author}{\bibinfo{person}{Odd~Erik Gundersen} {and} \bibinfo{person}{Sigbj{\o}rn Kjensmo}.} \bibinfo{year}{2018}\natexlab{}.
\newblock \showarticletitle{State of the Art: Reproducibility in Artificial Intelligence}. In \bibinfo{booktitle}{\emph{Proceedings of the AAAI Conference on Artificial Intelligence}}, Vol.~\bibinfo{volume}{32}.
\newblock
\href{https://doi.org/10.1609/aaai.v32i1.11503}{doi:\nolinkurl{10.1609/aaai.v32i1.11503}}


\bibitem[Harper and Konstan(2015)]%
        {harper2015movielens}
\bibfield{author}{\bibinfo{person}{F.~Maxwell Harper} {and} \bibinfo{person}{Joseph~A. Konstan}.} \bibinfo{year}{2015}\natexlab{}.
\newblock \showarticletitle{The MovieLens Datasets: History and Context}.
\newblock \bibinfo{journal}{\emph{ACM Transactions on Interactive Intelligent Systems}} \bibinfo{volume}{5}, \bibinfo{number}{4}, Article \bibinfo{articleno}{19} (\bibinfo{date}{Dec.} \bibinfo{year}{2015}).
\newblock
\showISSN{2160-6455}
\href{https://doi.org/10.1145/2827872}{doi:\nolinkurl{10.1145/2827872}}


\bibitem[J\"{a}rvelin and Kek\"{a}l\"{a}inen(2002)]%
        {jarvelinndcg2002}
\bibfield{author}{\bibinfo{person}{Kalervo J\"{a}rvelin} {and} \bibinfo{person}{Jaana Kek\"{a}l\"{a}inen}.} \bibinfo{year}{2002}\natexlab{}.
\newblock \showarticletitle{Cumulated Gain-Based Evaluation of IR Techniques}.
\newblock \bibinfo{journal}{\emph{ACM Transactions on Information Systems}} \bibinfo{volume}{20}, \bibinfo{number}{4} (\bibinfo{date}{Oct.} \bibinfo{year}{2002}), \bibinfo{pages}{422–446}.
\newblock
\showISSN{1046-8188}
\href{https://doi.org/10.1145/582415.582418}{doi:\nolinkurl{10.1145/582415.582418}}


\bibitem[Jeunen(2019)]%
        {jeunenRevisitingOfflineEvaluation2019}
\bibfield{author}{\bibinfo{person}{Olivier Jeunen}.} \bibinfo{year}{2019}\natexlab{}.
\newblock \showarticletitle{Revisiting Offline Evaluation for Implicit-Feedback Recommender Systems}. In \bibinfo{booktitle}{\emph{Proceedings of the 13th {{ACM Conference}} on {{Recommender Systems}}}} \emph{(\bibinfo{series}{{{RecSys}} '19})}. \bibinfo{pages}{596--600}.
\newblock
\showISBNx{978-1-4503-6243-6}
\href{https://doi.org/10.1145/3298689.3347069}{doi:\nolinkurl{10.1145/3298689.3347069}}


\bibitem[Lathia et~al\mbox{.}(2009)]%
        {lathiaEvaluatingCollaborativeFiltering2009}
\bibfield{author}{\bibinfo{person}{Neal Lathia}, \bibinfo{person}{Stephen Hailes}, {and} \bibinfo{person}{Licia Capra}.} \bibinfo{year}{2009}\natexlab{}.
\newblock \showarticletitle{Evaluating {{Collaborative Filtering Over Time}}}. In \bibinfo{booktitle}{\emph{{{SIGIR}} '09 {{Workshop}} on the {{Future}} of {{IR Evaluation}}}}.
\newblock
\urldef\tempurl%
\url{http://www.cs.ucl.ac.uk/staff/n.lathia/papers/lathia_ireval09.pdf}
\showURL{%
\tempurl}


\bibitem[Meng et~al\mbox{.}(2020)]%
        {meng2020exploring}
\bibfield{author}{\bibinfo{person}{Zaiqiao Meng}, \bibinfo{person}{Richard McCreadie}, \bibinfo{person}{Craig Macdonald}, {and} \bibinfo{person}{Iadh Ounis}.} \bibinfo{year}{2020}\natexlab{}.
\newblock \showarticletitle{Exploring Data Splitting Strategies for the Evaluation of Recommendation Models}. In \bibinfo{booktitle}{\emph{Proceedings of the 14th ACM Conference on Recommender Systems}}. \bibinfo{pages}{681--686}.
\newblock
\href{https://doi.org/10.1145/3383313.3418479}{doi:\nolinkurl{10.1145/3383313.3418479}}


\bibitem[Mishra et~al\mbox{.}(2019)]%
        {mishra2019big}
\bibfield{author}{\bibinfo{person}{Monika Mishra}, \bibinfo{person}{Jaydeep Chopde}, \bibinfo{person}{Maitri Shah}, \bibinfo{person}{Pankti Parikh}, \bibinfo{person}{Rakshith~Chandan Babu}, {and} \bibinfo{person}{Jongwook Woo}.} \bibinfo{year}{2019}\natexlab{}.
\newblock \showarticletitle{Big Data Predictive Analysis of Amazon Product Review}. In \bibinfo{booktitle}{\emph{KSII The 14th Asia Pacific International Conference on Information Science and Technology, APIC-IST, KSII, Beijing, China}}.
\newblock
\urldef\tempurl%
\url{https://www.calstatela.edu/sites/default/files/amazonprodreviewapic-ist2019.pdf}
\showURL{%
\tempurl}


\bibitem[Moffat and Zobel(2008)]%
        {MoffatZobel2008}
\bibfield{author}{\bibinfo{person}{Alistair Moffat} {and} \bibinfo{person}{Justin Zobel}.} \bibinfo{year}{2008}\natexlab{}.
\newblock \showarticletitle{Rank-Biased Precision for Measurement of Retrieval Effectiveness}.
\newblock \bibinfo{journal}{\emph{ACM Transactions on Information Systems}} \bibinfo{volume}{27}, \bibinfo{number}{1}, Article \bibinfo{articleno}{2} (\bibinfo{date}{Dec.} \bibinfo{year}{2008}).
\newblock
\showISSN{1046-8188}
\href{https://doi.org/10.1145/1416950.1416952}{doi:\nolinkurl{10.1145/1416950.1416952}}


\bibitem[Nesvijevskaia(2021)]%
        {nesvijevskaia2021databook}
\bibfield{author}{\bibinfo{person}{Anna Nesvijevskaia}.} \bibinfo{year}{2021}\natexlab{}.
\newblock \showarticletitle{DATABOOK: A Standardised Framework for Dynamic Documentation of Algorithm Design During Data Science Projects}.
\newblock \bibinfo{journal}{\emph{IASSIST Quarterly}} \bibinfo{volume}{45}, \bibinfo{number}{2} (\bibinfo{year}{2021}).
\newblock
\href{https://doi.org/10.29173/iq989}{doi:\nolinkurl{10.29173/iq989}}


\bibitem[Nguyen et~al\mbox{.}(2014)]%
        {nguyenExploringFilterBubble2014}
\bibfield{author}{\bibinfo{person}{Tien~T Nguyen}, \bibinfo{person}{Pik-Mai Hui}, \bibinfo{person}{F~Maxwell Harper}, \bibinfo{person}{Loren Terveen}, {and} \bibinfo{person}{Joseph~A Konstan}.} \bibinfo{year}{2014}\natexlab{}.
\newblock \showarticletitle{Exploring the {{Filter Bubble}}: {{The Effect}} of {{Using Recommender Systems}} on {{Content Diversity}}}. In \bibinfo{booktitle}{\emph{Proceedings of the 23rd {{International Conference}} on {{World Wide Web}}}} \emph{(\bibinfo{series}{{{WWW}} '14})}. \bibinfo{pages}{677--686}.
\newblock
\href{https://doi.org/10.1145/2566486.2568012}{doi:\nolinkurl{10.1145/2566486.2568012}}


\bibitem[Olteanu et~al\mbox{.}(2019)]%
        {olteanu2019social}
\bibfield{author}{\bibinfo{person}{Alexandra Olteanu}, \bibinfo{person}{Carlos Castillo}, \bibinfo{person}{Fernando Diaz}, {and} \bibinfo{person}{Emre K{\i}c{\i}man}.} \bibinfo{year}{2019}\natexlab{}.
\newblock \showarticletitle{Social Data: Biases, Methodological Pitfalls, and Ethical Boundaries}.
\newblock \bibinfo{journal}{\emph{Frontiers in Big Data}}  \bibinfo{volume}{2} (\bibinfo{year}{2019}), \bibinfo{pages}{13}.
\newblock
\href{https://doi.org/10.3389/fdata.2019.00013}{doi:\nolinkurl{10.3389/fdata.2019.00013}}


\bibitem[Pagano et~al\mbox{.}(2023)]%
        {pagano2023bias}
\bibfield{author}{\bibinfo{person}{Tiago~P Pagano}, \bibinfo{person}{Rafael~B Loureiro}, \bibinfo{person}{Fernanda~VN Lisboa}, \bibinfo{person}{Rodrigo~M Peixoto}, \bibinfo{person}{Guilherme~AS Guimar{\~a}es}, \bibinfo{person}{Gustavo~OR Cruz}, \bibinfo{person}{Maira~M Araujo}, \bibinfo{person}{Lucas~L Santos}, \bibinfo{person}{Marco~AS Cruz}, \bibinfo{person}{Ewerton~LS Oliveira}, {et~al\mbox{.}}} \bibinfo{year}{2023}\natexlab{}.
\newblock \showarticletitle{Bias and Unfairness in Machine Learning Models: A Systematic Review on Datasets, Tools, Fairness Metrics, and Identification and Mitigation Methods}.
\newblock \bibinfo{journal}{\emph{Big Data and Cognitive Computing}} \bibinfo{volume}{7}, \bibinfo{number}{1} (\bibinfo{year}{2023}), \bibinfo{pages}{15}.
\newblock
\href{https://doi.org/10.3390/bdcc7010015}{doi:\nolinkurl{10.3390/bdcc7010015}}


\bibitem[Pushkarna et~al\mbox{.}(2022)]%
        {pushkarna2022datacards}
\bibfield{author}{\bibinfo{person}{Mahima Pushkarna}, \bibinfo{person}{Andrew Zaldivar}, {and} \bibinfo{person}{Oddur Kjartansson}.} \bibinfo{year}{2022}\natexlab{}.
\newblock \showarticletitle{Data Cards: Purposeful and Transparent Dataset Documentation for Responsible AI}. In \bibinfo{booktitle}{\emph{Proceedings of the 2022 ACM Conference on Fairness, Accountability, and Transparency}} \emph{(\bibinfo{series}{FAccT '22})}. \bibinfo{pages}{1776–1826}.
\newblock
\showISBNx{9781450393522}
\href{https://doi.org/10.1145/3531146.3533231}{doi:\nolinkurl{10.1145/3531146.3533231}}


\bibitem[Raj and Ekstrand(2023)]%
        {raj2023unified}
\bibfield{author}{\bibinfo{person}{Amifa Raj} {and} \bibinfo{person}{Michael Ekstrand}.} \bibinfo{year}{2023}\natexlab{}.
\newblock \showarticletitle{Unified Browsing Models for Linear and Grid Layouts}.
\newblock \bibinfo{journal}{\emph{arXiv preprint arXiv:2310.12524}} (\bibinfo{year}{2023}).
\newblock
\href{https://doi.org/10.48550/arXiv.2310.12524}{doi:\nolinkurl{10.48550/arXiv.2310.12524}}


\bibitem[Raj and Ekstrand(2024)]%
        {gridlayout2024amifaraj}
\bibfield{author}{\bibinfo{person}{Amifa Raj} {and} \bibinfo{person}{Michael~D. Ekstrand}.} \bibinfo{year}{2024}\natexlab{}.
\newblock \showarticletitle{Towards Optimizing Ranking in Grid-Layout for Provider-Side Fairness}. In \bibinfo{booktitle}{\emph{Advances in Information Retrieval}} (2024), \bibfield{editor}{\bibinfo{person}{Nazli Goharian}, \bibinfo{person}{Nicola Tonellotto}, \bibinfo{person}{Yulan He}, \bibinfo{person}{Aldo Lipani}, \bibinfo{person}{Graham McDonald}, \bibinfo{person}{Craig Macdonald}, {and} \bibinfo{person}{Iadh Ounis}} (Eds.). \bibinfo{pages}{90--105}.
\newblock
\showISBNx{978-3-031-56069-9}
\href{https://doi.org/10.1007/978-3-031-56069-9_7}{doi:\nolinkurl{10.1007/978-3-031-56069-9_7}}


\bibitem[Rendle et~al\mbox{.}(2009)]%
        {bpr2009}
\bibfield{author}{\bibinfo{person}{Steffen Rendle}, \bibinfo{person}{Christoph Freudenthaler}, \bibinfo{person}{Zeno Gantner}, {and} \bibinfo{person}{Lars Schmidt-Thieme}.} \bibinfo{year}{2009}\natexlab{}.
\newblock \showarticletitle{BPR: Bayesian Personalized Ranking from Implicit Feedback}. In \bibinfo{booktitle}{\emph{Proceedings of the Twenty-Fifth Conference on Uncertainty in Artificial Intelligence}} \emph{(\bibinfo{series}{UAI '09})}. \bibinfo{pages}{452–461}.
\newblock
\showISBNx{9780974903958}


\bibitem[Takács et~al\mbox{.}(2011)]%
        {takacsApplicationsConjugateGradient2011}
\bibfield{author}{\bibinfo{person}{Gábor Takács}, \bibinfo{person}{István Pilászy}, {and} \bibinfo{person}{Domonkos Tikk}.} \bibinfo{year}{2011}\natexlab{}.
\newblock \showarticletitle{Applications of the Conjugate Gradient Method for Implicit Feedback Collaborative Filtering}. In \bibinfo{booktitle}{\emph{Proceedings of the Fifth {{ACM}} Conference on {{Recommender}} Systems}} (2011-10-23). \bibinfo{pages}{297--300}.
\newblock
\showISBNx{978-1-4503-0683-6}
\href{https://doi.org/10.1145/2043932.2043987}{doi:\nolinkurl{10.1145/2043932.2043987}}


\bibitem[Thelwall and Kousha(2017)]%
        {thelwall2017goodreads}
\bibfield{author}{\bibinfo{person}{Mike Thelwall} {and} \bibinfo{person}{Kayvan Kousha}.} \bibinfo{year}{2017}\natexlab{}.
\newblock \showarticletitle{Goodreads: A Social Network Site for Book Readers}.
\newblock \bibinfo{journal}{\emph{Journal of the Association for Information Science and Technology}} \bibinfo{volume}{68}, \bibinfo{number}{4} (\bibinfo{year}{2017}), \bibinfo{pages}{972--983}.
\newblock
\href{https://doi.org/10.1002/asi.23733}{doi:\nolinkurl{10.1002/asi.23733}}


\bibitem[Verboord(2011)]%
        {Verboord2011}
\bibfield{author}{\bibinfo{person}{Marc Verboord}.} \bibinfo{year}{2011}\natexlab{}.
\newblock \showarticletitle{Cultural Products Go Online: {{Comparing}} the Internet and Print Media on Distributions of Gender, Genre and Commercial Success}.
\newblock \bibinfo{journal}{\emph{Communications}} \bibinfo{volume}{36}, \bibinfo{number}{4} (\bibinfo{year}{2011}), \bibinfo{pages}{441--462}.
\newblock
\href{https://doi.org/doi:10.1515/comm.2011.022}{doi:\nolinkurl{doi:10.1515/comm.2011.022}}


\bibitem[Wan and McAuley(2018)]%
        {wan2018item}
\bibfield{author}{\bibinfo{person}{Mengting Wan} {and} \bibinfo{person}{Julian~J. McAuley}.} \bibinfo{year}{2018}\natexlab{}.
\newblock \showarticletitle{Item Recommendation on Monotonic Behavior Chains}. In \bibinfo{booktitle}{\emph{Proceedings of the 12th ACM Conference on Recommender Systems}}. \bibinfo{pages}{86--94}.
\newblock
\href{https://doi.org/10.1145/3240323.3240369}{doi:\nolinkurl{10.1145/3240323.3240369}}


\bibitem[Wan et~al\mbox{.}(2019)]%
        {wan2019fine}
\bibfield{author}{\bibinfo{person}{Mengting Wan}, \bibinfo{person}{Rishabh Misra}, \bibinfo{person}{Ndapa Nakashole}, {and} \bibinfo{person}{Julian~J. McAuley}.} \bibinfo{year}{2019}\natexlab{}.
\newblock \showarticletitle{Fine-Grained Spoiler Detection from Large-Scale Review Corpora}. In \bibinfo{booktitle}{\emph{Proceedings of the 57th Conference of the Association for Computational Linguistics}}. \bibinfo{pages}{2605--2610}.
\newblock
\href{https://doi.org/10.18653/v1/P19-1248}{doi:\nolinkurl{10.18653/v1/P19-1248}}


\bibitem[Wang et~al\mbox{.}(2023)]%
        {wang2023surveyfairness}
\bibfield{author}{\bibinfo{person}{Yifan Wang}, \bibinfo{person}{Weizhi Ma}, \bibinfo{person}{Min Zhang}, \bibinfo{person}{Yiqun Liu}, {and} \bibinfo{person}{Shaoping Ma}.} \bibinfo{year}{2023}\natexlab{}.
\newblock \showarticletitle{A Survey on the Fairness of Recommender Systems}.
\newblock \bibinfo{journal}{\emph{ACM Transactions on Information Systems}} \bibinfo{volume}{41}, \bibinfo{number}{3}, Article \bibinfo{articleno}{52} (\bibinfo{date}{Feb.} \bibinfo{year}{2023}).
\newblock
\showISSN{1046-8188}
\href{https://doi.org/10.1145/3547333}{doi:\nolinkurl{10.1145/3547333}}


\bibitem[Wang et~al\mbox{.}(2022)]%
        {wang2022surrogate}
\bibfield{author}{\bibinfo{person}{Yuyan Wang}, \bibinfo{person}{Mohit Sharma}, \bibinfo{person}{Can Xu}, \bibinfo{person}{Sriraj Badam}, \bibinfo{person}{Qian Sun}, \bibinfo{person}{Lee Richardson}, \bibinfo{person}{Lisa Chung}, \bibinfo{person}{Ed~H. Chi}, {and} \bibinfo{person}{Minmin Chen}.} \bibinfo{year}{2022}\natexlab{}.
\newblock \showarticletitle{Surrogate for Long-Term User Experience in Recommender Systems}. In \bibinfo{booktitle}{\emph{Proceedings of the 28th ACM SIGKDD Conference on Knowledge Discovery and Data Mining}}. \bibinfo{pages}{4100--4109}.
\newblock
\href{https://doi.org/10.1145/3534678.3539073}{doi:\nolinkurl{10.1145/3534678.3539073}}


\bibitem[Yalcin and Bilge(2022)]%
        {yalcin2022evaluating}
\bibfield{author}{\bibinfo{person}{Emre Yalcin} {and} \bibinfo{person}{Alper Bilge}.} \bibinfo{year}{2022}\natexlab{}.
\newblock \showarticletitle{Evaluating Unfairness of Popularity Bias in Recommender Systems: A Comprehensive User-Centric Analysis}.
\newblock \bibinfo{journal}{\emph{Information Processing \& Management}} \bibinfo{volume}{59}, \bibinfo{number}{6} (\bibinfo{year}{2022}).
\newblock
\href{https://doi.org/10.1016/j.ipm.2022.103100}{doi:\nolinkurl{10.1016/j.ipm.2022.103100}}


\end{thebibliography}

\end{document}